\documentclass[journal]{IEEEtran}

\usepackage{amssymb}
\usepackage[cmex10]{amsmath}
\usepackage{stfloats}
\usepackage{graphicx}
\usepackage{subfigure}
\usepackage{tabularx}
\usepackage{epsfig,epsf,color,balance,cite}
\usepackage{verbatim}
\usepackage{url}
\usepackage{bm}

\newtheorem{theorem}{\bf Theorem}

\newtheorem{lemma}{\bf Lemma}

\usepackage{algorithm}
\usepackage{algorithmic}

\usepackage{color}
\definecolor{myc1}{rgb}{0,0,0}

\definecolor{myc2}{rgb}{0,0,0}

\definecolor{myc3}{rgb}{0,0,0}

\definecolor{myc5}{rgb}{0,0,0}

\begin{document}

\title{Energy Efficient Federated Learning Over Wireless Communication Networks}

\author{
\IEEEauthorblockN{Zhaohui Yang,
                  Mingzhe Chen,
                  Walid Saad, \IEEEmembership{Fellow, IEEE},
                  Choong Seon Hong, \IEEEmembership{Senior Member, IEEE},
                  and Mohammad Shikh-Bahaei, \IEEEmembership{Senior Member, IEEE}
                   \vspace{-2em}
                  }
 \thanks {This work was supported in part by the EPSRC through the Scalable Full Duplex Dense Wireless Networks (SENSE) grant EP/P003486/1, and by the U.S. National Science Foundation under Grant CNS-1814477. A preliminary version of this work was by IEEE ICML 2020  \cite{yang2019eeFLConf}.}
\thanks{Z. Yang  and M. Shikh-Bahaei are with the Centre for Telecommunications Research, Department of Engineering, King's College London, WC2R 2LS, UK, Emails: yang.zhaohui@kcl.ac.uk, m.sbahaei@kcl.ac.uk.}
\thanks{M. Chen is with the Electrical Engineering Department of Princeton University, NJ, 08544, USA, and also with the Chinese University of Hong Kong, Shenzhen, 518172, China, Email: mingzhec@princeton.edu.}
\thanks{W. Saad is with Wireless@VT, Bradley Department of Electrical and Computer Engineering, Virginia Tech, Blacksburg, VA, 24060, USA, Email: walids@vt.edu.}
\thanks{C. Hong is with the Department of Computer Science and Engineering, Kyung Hee University, Yongin-si, Gyeonggi-do 17104, Rep. of Korea, Email: cshong@khu.ac.kr.}
 }

\maketitle

\begin{abstract}
In this paper, the problem of energy efficient transmission and computation resource allocation for federated learning (FL) over wireless communication networks is investigated. In the considered model, each user exploits limited local computational resources to train a local FL  model with its collected data and, then, sends the trained FL model to a base station (BS) which aggregates the local FL model and broadcasts it back to all of the users. Since FL involves an exchange of a learning model between users and the BS, both computation and communication latencies are determined by the learning accuracy level. Meanwhile, due to the limited energy budget of the wireless users, both local computation energy and transmission energy must be considered during the FL process. This joint learning and communication problem is formulated as an optimization problem whose goal is to minimize the total energy consumption of the system under a latency constraint. To solve this problem, an iterative algorithm is proposed where, at every step,  closed-form solutions for time allocation, bandwidth allocation, power control, computation frequency, and learning accuracy are derived. Since the iterative algorithm requires an initial feasible solution, we construct the completion time minimization problem and a bisection-based algorithm is proposed to obtain the optimal solution, which is a feasible solution to the original energy minimization problem. Numerical results show that the proposed algorithms can reduce up to  59.5\% energy consumption  compared to the conventional FL method.
\end{abstract}

\IEEEpeerreviewmaketitle

\vspace{-0.5em}
\section{Introduction}\vspace{-.5em}
In future wireless systems, due to privacy constraints and limited communication resources for data transmission, it is impractical for all wireless devices to transmit all of their collected data to a data center that can use the collected data to implement centralized machine learning algorithms for data analysis and inference \cite{wang2018edge,sun2019application,liu2019machine,bonawitz2019towards,8755300,chen2019joint}.
{\color{myc1}{To this end, distributed learning frameworks are needed, to enable the wireless devices to collaboratively build a shared learning model with training their collected data locally \cite{saad2019vision,Wireless2018Park,chen2018federated,tembine2018distributed,samarakoon2018distributed,8839651,8755300,abad2019hierarchical,nishio2019client,karimireddy2019scaffold}.
One of the most promising distributed learning algorithms is the emerging federated learning (FL) framework that will be adopted in future Internet of Things (IoT) systems \cite{mcmahan2016communication,smith2017federated,8851249,8664630,DBLP:journals/corr/abs-1907-06426,ahn2019wireless,7875131,8038869,8379427,long2020reflections}.
In FL, wireless devices can cooperatively execute a learning task by only uploading local learning model to the base station (BS) instead of sharing the entirety of their training data \cite{konevcny2016federated}.}}
{\color{myc3}{Using gradient sparsification, a digital transmission scheme based on gradient quantization was investigated in \cite{amiri2019machine}.}}
To implement FL over wireless networks, the wireless devices must transmit their local training results over wireless links \cite{zhu2018towards}, which can affect  the performance of FL due to limited wireless resources (such as time and bandwidth).
In addition, the limited energy of wireless devices is a key challenge for deploying FL. Indeed, because of these resource constraints, it is necessary to optimize the energy efficiency for FL implementation.



Some of the challenges of FL over wireless networks have been studied in \cite{zhu2018low,Vu2019CellFreeMM,sun2019energyaware,
Yang2018FLOA,Zeng2019EEFL,chen2019joint,tran2019federated}.
To minimize latency, a broadband analog aggregation multi-access scheme was designed in \cite{zhu2018low} for FL by exploiting the
waveform-superposition property of a multi-access channel.
{\color{myc2} {An FL training minimization problem was investigated in \cite{Vu2019CellFreeMM} for cell-free massive multiple-input multiple-output (MIMO) systems.}}
{\color{myc2}{For FL with redundant data, an energy-aware user scheduling policy was proposed in \cite{sun2019energyaware} to maximize the average number of scheduled users.}}
To improve the statistical learning performance for on-device distributed training, the authors in \cite{Yang2018FLOA} developed a novel sparse and low-rank modeling approach. 
The work in \cite{Zeng2019EEFL} introduced
an energy-efficient strategy for bandwidth allocation under learning performance constraints.
However, the works in \cite{zhu2018low,Vu2019CellFreeMM,sun2019energyaware,
Yang2018FLOA,Zeng2019EEFL} focused on the {\color{myc1}{delay/energy  for wireless transmission}} without considering the delay/energy tradeoff between learning and transmission.
Recently, the works in \cite{chen2019joint} and \cite{tran2019federated} considered both local learning and wireless transmission energy.
{\color{myc1}{In \cite{chen2019joint}, we investigated the FL loss function minimization problem with taking into account packet errors over wireless links.
However, this prior work ignored the computation delay of local FL model. 
The authors in \cite{tran2019federated} considered the sum computation and transmission energy minimization problem for FL.
However, the solution in \cite{tran2019federated} requires all users to upload their learning model synchronously.
Meanwhile, the work in \cite{tran2019federated} did not provide any convergence analysis for FL.}} 

The main contribution of this paper is a novel energy efficient computation and transmission resource allocation scheme for FL over wireless communication networks.  Our key contributions include:
\begin{itemize}
\item
We study the performance of FL algorithm over wireless communication networks for a scenario in which each user locally computes its FL model under a given learning accuracy and the BS  broadcasts the aggregated FL model to all users.
For the considered FL algorithm, we first derive the  convergence rate.
%
\item 
We formulate a joint computation and transmission optimization problem aiming to minimize the total energy consumption for local computation and wireless transmission. To solve this problem, an iterative algorithm is proposed with low complexity. At each step of this algorithm, we derive new closed-form solutions for the time allocation, bandwidth allocation, power control, computation frequency, and learning accuracy.
\item {\color{myc1}{To obtain a feasible solution for the total energy minimization problem, we construct the FL completion time minimization problem.}} We theoretically
show that the completion time is a convex function of the learning accuracy. 
Based on this theoretical finding, we propose a bisection-based algorithm to obtain the optimal solution for the FL completion time minimization.
\item Simulation results show that the proposed scheme that jointly considers computation and transmission optimization can achieve up to 59.5\%  energy reduction compared to the conventional FL method.
\end{itemize}


The rest of this paper is organized as follows.
The system model is described in Section \uppercase\expandafter{\romannumeral2}.
Section \uppercase\expandafter{\romannumeral3} provides problem formulation and
the resource allocation for total energy minimization.
The algorithm to find a feasible solution of the original energy minimization problem is given in Section \uppercase\expandafter{\romannumeral4}.
Simulation results are analyzed in Section \uppercase\expandafter{\romannumeral5}. Conclusions are drawn in Section \uppercase\expandafter{\romannumeral6}.

\vspace{-0.5em}
\section{System Model}\vspace{-0.5em}

\begin{figure}
\centering
\includegraphics[width=3.0in]{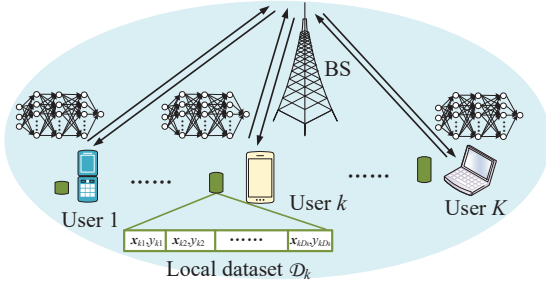}
\vspace{-1em}
\caption{Illustration of the considered model for FL over wireless communication networks.}
\vspace{-1em}
\label{sys}
\end{figure}

Consider a cellular network that consists of one BS serving  $K$ users, as shown in Fig.~\ref{sys}.
Each user $k$ has a local dataset $\mathcal D_k$ with $D_k$ data samples.
For each dataset $\mathcal D_k=\{\boldsymbol x_{kl},y_{kl}\}_{l=1}^{D_k}$, $\boldsymbol x_{kl}\in\mathbb R^d$ is an input vector of user $k$ and $y_{kl}$ is its corresponding output\footnote{For simplicity, we consider
an FL algorithm with a single output. In future work, our approach will be extended  to the
case with multiple outputs.}.
The BS and all users
cooperatively perform
an FL algorithm over wireless networks for data analysis and inference.
 Hereinafter, the FL model that is trained by each user's dataset is called the \emph{local FL model}, while the FL model that is generated by the BS using local FL model inputs from all users is called the \emph{global FL model}.

\vspace{-0.5em}
\subsection{{\color{myc3}Computation and Transmission Model}}
\vspace{-0.5em}


\begin{figure}
\centering
\includegraphics[width=3.5in]{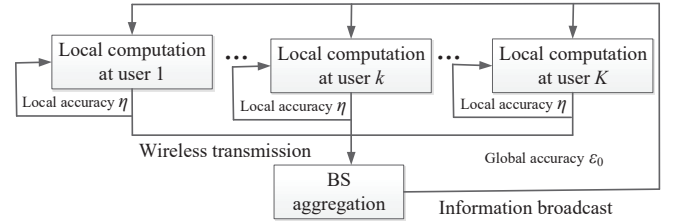}
\vspace{-1em}
\caption{The FL procedure between users and the BS.}\label{sys1}
\vspace{-1em}
\end{figure}

The FL procedure between the users and their serving BS is shown in Fig.~\ref{sys1}.
From this figure,
the FL procedure contains three steps at each  iteration: local computation at each user (using several local iterations), local FL model transmission for each user, and result aggregation and broadcast at the BS.
The local computation step is essentially the phase during which each user calculates its local FL model by using its local data set and the received global FL model.
\subsubsection{Local Computation}

Let $f_k$ be the computation capacity of user $k$, which is measured by the number of CPU cycles per second.
The computation time at user $k$ needed for data processing is:
\vspace{-0.5em}
\begin{equation}\label{sys1eq5_2}\vspace{-0.5em}
\tau_{k} =\frac{I_k C_k D_k  }{f_k}, \quad \forall k\in\mathcal K,
\end{equation}
where $C_{k}$ (cycles/sample) is the number of CPU cycles required for computing one sample data at user $k$ and $I_k$ is the number of local iterations at user $k$. 
{\color{myc3}According to Lemma 1 in \cite{7572018}, the energy consumption for computing a total number of $C_kD_k$ CPU cycles at user $k$ is:
\vspace{-0.5em}
\begin{equation}\vspace{-0.5em}
{\color{myc5}E_{k1}^{\text C}}=\kappa C_k D_k  f_k^{2},
\end{equation}
where  $\kappa$ is the effective switched capacitance that depends on the chip architecture.
To compute the local FL model, user $k$ needs to compute $C_kD_k$ CPU cycles with $I_k$ local iterations, which means that the total computation energy at user $k$ is:
\vspace{-.5em}
\begin{equation}\label{sys1eq6}\vspace{-0.5em}
E_{k}^{\text{C}}=I_kE_{k1}^{\text C}=\kappa I_k C_k D_k   f_k^{2}.
\end{equation}
}



\subsubsection{Wireless Transmission}\vspace{-.5em}
After local computation, all users upload their local FL model to the BS via frequency domain multiple access (FDMA).
The achievable rate of user $k$ can be given by:
\vspace{-.5em}
\begin{equation}\label{sys1eq7}\vspace{-0.5em}
r_k=b_k\log_2\left(1+ \frac{g_kp_k}{N_0b_k}
\right), \quad \forall k\in\mathcal K,
\end{equation}
where $b_k$ is the bandwidth allocated to user $k$, $p_k$ is the average transmit power of user $k$, $g_k$ is the channel gain between user $k$ and the BS, and $N_0$ is the power spectral density of the Gaussian noise.
{\color{myc3}Note that the Shannon capacity \eqref{sys1eq7} serves as an
upper bound of the transmission rate.}
Due to limited bandwidth of the system, we have:
$\sum_{k=1}^K b_k \leq B$,
where $B$ is the total bandwidth.

In this step, user $k$ needs to upload the local FL model 
 to the BS.
Since the dimensions of the local FL model are fixed for all users, the data size that each user needs to upload is constant, and can be denoted by $s$.
To upload data of size $s$ within  transmission time $t_k$, we must have:
$t_k{r_k} \geq s$.
To transmit data of size $s$ within a time duration $t_k$, the wireless transmit energy of user $k$ will be:
$E_{k}^{\text{T}} = t_kp_k$.

\subsubsection{Information Broadcast}
In this step, the BS aggregates the global FL model.
The BS broadcasts the global FL model to all users in the downlink.
Due to the high transmit power at the BS and the high bandwidth that can be used for data broadcasting,
the downlink time is neglected compared to the uplink data transmission time.
It can be observed that the local data $\mathcal D_k$ is not accessed by the BS, so as to protect the privacy of users, as is required by FL.


According to the above FL model, the energy consumption of each user includes both local computation energy $E_{k}^{\text{C}}$ and wireless transmission energy $E_{k}^{\text{T}}$.
Denote the number of global iterations by $I_0$, and the total energy consumption of all users that participate in FL will be:
\vspace{-.5em}
\begin{align}\label{sys1energy0}\vspace{-.5em}
E&=I_0 \sum_{k=1}^K (E_{k}^{\text{C}}+E_{k}^{\text{T}}). 
\end{align}

%

\begin{figure}
\centering
\includegraphics[width=3.0in]{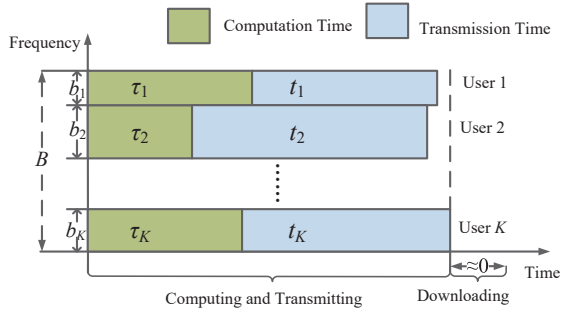}
\vspace{-1em}
\caption{{\color{myc1}{An implementation for the FL algorithm via FDMA.}}}\label{sys2}
\vspace{-1em}
\end{figure}

{\color{myc1}{Hereinafter, the total time needed for completing the execution of the FL algorithm is called \emph{completion time}.}}
The completion time of each user includes the local computation time and transmission time, as shown in Fig.~\ref{sys2}.
Based on \eqref{sys1eq5_2}, the completion time $T_k$ of user $k$ will be:
\vspace{-.5em}
\begin{align}\vspace{-.5em}\label{Timeconseq1}
T_k=I_0(\tau_k+t_k)
=I_0\left(\frac{I_kC_kD_k}{f_k}+t_k
\right).
\end{align}
Let $T$ be the maximum completion time for training the entire FL algorithm and we have:
\vspace{-.5em}
\begin{align}\label{sys1time0}\vspace{-.5em}
T_k \leq T, \quad \forall k\in\mathcal K.
\end{align}


\vspace{-1em}
\subsection{{\color{myc3}FL Model}}
\vspace{-0.5em}

{\color{myc1}{We define a vector $\boldsymbol w$ to capture the parameters related to the global FL  model.}}
We introduce the loss function $f(\boldsymbol w,\boldsymbol x_{kl}, y_{kl})$, that captures the  FL performance over input vector $\boldsymbol{x}_{kl}$ and output $y_{kl}$.
For different learning tasks, the loss function will be different.
For example, $f(\boldsymbol w,\boldsymbol x_{kl}, y_{kl})=\frac1 2 (\boldsymbol x_{kl}^T\boldsymbol w-y_{kl})^2$ for linear regression and
$f(\boldsymbol w,\boldsymbol x_{kl}, y_{kl})=-\log (1+\exp(-y_{kl}\boldsymbol x_{kl}^T\boldsymbol w))$ for logistic regression.
Since the dataset of user $k$ is $\mathcal D_k$, the total loss function of user $k$ will be:
\vspace{-.5em}
\begin{equation}\label{sys0eq1}\vspace{-.5em}
F_k(\boldsymbol w,  x_{k1},y_{k1}, \cdots, x_{kD_k},y_{kD_k} )=\frac 1 {D_k} \sum_{l=1}^{D_k} f(\boldsymbol w,\boldsymbol x_{kl}, y_{kl}).
\end{equation}
{\color{myc2}{Note that function $f(\boldsymbol w,\boldsymbol x_{kl}, y_{kl})$ is the loss function of user $k$ with one data sample and function $F_k(\boldsymbol w,  x_{k1},y_{k1}, \cdots, x_{kD_k},y_{kD_k} )$ is the total loss function of user $k$ with the whole local dataset. In the following, $F_k(\boldsymbol w,  x_{k1},y_{k1}, \cdots, x_{kD_k},y_{kD_k} )$ is denoted by $F_k(\boldsymbol w)$ for simplicity of notation.}}

In order to deploy an FL algorithm, it is necessary to train the underlying model. Training is done in order to generate a unified FL model for all users without sharing any datasets.
The FL training problem can be formulated as \cite{konevcny2016federated,wang2018edge,8664630}:
\vspace{-.5em}
\begin{equation}\label{sys0eq2}\vspace{-.5em}
\min_{\boldsymbol w} F(\boldsymbol w)\triangleq\sum_{k=1}^K \frac{D_k}{D} F_k(\boldsymbol w)=\frac{1}{D}\sum_{k=1}^K\sum_{l=1}^{D_k} f(\boldsymbol w,\boldsymbol x_{kl}, y_{kl}),
\end{equation}
where $D=\sum_{k=1}^KD_k$ is the total data samples of all users.

{\color{myc3}To solve problem \eqref{sys0eq2}, we adopt the distributed approximate Newton (DANE) algorithm in \cite{konevcny2016federated}, which is summarized in Algorithm~1. Note that the gradient descent (GD) is used at each user in Algorithm 1.
 One can use stochastic gradient descent (SGD) to decrease the computation complexity for cases in which a relatively low accuracy can be tolerated.
 However, if high accuracy was needed, gradient descent (GD) would be preferred  \cite{konevcny2016federated}.
 Moreover, the number of global iterations is higher in SGD than in GD.
Due to limited wireless resources, SGD may not be efficient since SGD requires more iterations for wireless transmissions compared to GD.
The  DANE algorithm is designed to  solve a general {\color{myc5}local optimization problem}, before averaging the solutions of all users. The DANE algorithm relies on the similarity of the Hessians of local objectives, representing their iterations as an average of inexact Newton steps.}
\begin{algorithm}[t]
\caption{FL Algorithm}
\begin{algorithmic}[1]
 \STATE Initialize 
 global model $\boldsymbol w^0$ and global iteration number $n=0$.
 \REPEAT
 \STATE Each user $k$ computes $\nabla  F_k(\boldsymbol w^{(n)})$ and sends it to the BS.
 \STATE The BS computes $ \nabla  F(\boldsymbol w^{(n)})= \frac 1 K\sum_{k=1}^K \nabla  F_k(\boldsymbol w^{(n)})$,
 which is broadcast to all users.
 \STATE  \textbf{parallel for} user $k\in\mathcal K=\{1,\cdots, K\}$
\STATE \quad  Initialize {\color{myc3}the local iteration number $i=0$} and set $\boldsymbol h_k^{(n),(0)}=\boldsymbol 0$.
 \STATE \quad \textbf{repeat}
\STATE \quad \quad Update $\boldsymbol h_k^{(n),(i+1)}=\boldsymbol h_k^{(n),(i)}- \delta\nabla G_k(\boldsymbol w^{(n)}, \boldsymbol h_k^{(n),(i)})$ and $i=i+1$.
 \STATE \quad \textbf{until}
{the accuracy $\eta$ of  {\color{myc5}local optimization problem}  \eqref{sys0eq3_1} is obtained.}
 \STATE \quad Denote $\boldsymbol h_k^{(n)}=\boldsymbol h_k^{(n),(i)}$ and each user sends $\boldsymbol h_k^{(n)}$ to the BS.
  \STATE \textbf{end for}
  \STATE   The BS computes
  \vspace{-.5em}
   \begin{equation}\label{sys0eq3_5}  \vspace{-.5em}
\boldsymbol w^{(n+1)}=\boldsymbol w^{(n)}+\frac {1} K \sum_{k=1}^K\boldsymbol h_k^{(n)},
 \end{equation}
 and broadcasts the value to all users.
 \STATE Set $n=n+1$.
 \UNTIL the accuracy $\epsilon_0$ of problem \eqref{sys0eq2} is obtained.
\end{algorithmic}
\end{algorithm}
In Algorithm 1, we can see that, at every FL iteration, each user downloads the global FL model from the BS for local computation, while  the BS periodically gathers the local FL model from all users and sends the updated global FL model back to all users.
{\color{myc3}According to Algorithm 1,
since the {\color{myc5}local optimization problem} is solved with several local iterations at each global iteration, the number of full gradient computations will be smaller than the number of local updates.}

We define $\boldsymbol w^{(n)}$ as the global FL model at a given iteration $n$.
{\color{myc3}In practice, each user {\color{myc5}solves} the {\color{myc5}local optimization problem}:
\vspace{-.5em}
\begin{align}\label{sys0eq3_1}  \vspace{-.5em}
  \min_{\boldsymbol h_k\in\mathbb R^d}\quad G_k&(\boldsymbol w^{(n)}, \boldsymbol h_k)\triangleq  F_k(\boldsymbol w^{(n)}+ \boldsymbol h_k)
  \nonumber \\&
-(\nabla  F_k(\boldsymbol w^{(n)})-  \xi \nabla  F(\boldsymbol w^{(n)}))^T \boldsymbol h_k.
\end{align}
 In problem \eqref{sys0eq3_1}, $\xi$ is a constant value.
Any solution $\boldsymbol h_k$ of problem \eqref{sys0eq3_1} represents the difference between the global FL model and  local FL model for user $k$, 
i.e., $\boldsymbol w^{(n)}+\boldsymbol h_k$ is the  local FL model of user $k$ at iteration $n$.
The {\color{myc5}local optimization problem} in \eqref{sys0eq3_1} depends only on the local data  and the gradient of the global loss function. The {\color{myc5}local optimization problem} is then solved, and the updates from individual users are averaged to form a new iteration. This approach allows any algorithm to be used to solve the {\color{myc5}local optimization problem}. Note that
the number of iterations needed to upload the local FL model (i.e., the number of global iterations) in Algorithm 1 is always smaller than in  the conventional FL algorithms that do not consider a {\color{myc5}local optimization problem} \eqref{sys0eq3_1} \cite{konevcny2016federated}.
To solve the {\color{myc5}local optimization problem} in \eqref{sys0eq3_1}, we use the gradient method:
\vspace{-.5em}
\begin{equation}\label{applemmacon2eq1}\vspace{-.5em}
\boldsymbol h_k^{(n),(i+1)}=\boldsymbol h_k^{(n),(i)}- \delta\nabla G_k(\boldsymbol w^{(n)}, \boldsymbol h_k^{(n),(i)}),
\end{equation}
where $\delta$ is the step size, $\boldsymbol h_k^{(n),(i)}$ is the value of $\boldsymbol h_k$
at the $i$-th local iteration with given vector $\boldsymbol w^{(n)}$, and $\nabla G_k(\boldsymbol w^{(n)}, \boldsymbol h_k^{(n),(i)})$ is the gradient of function $G_k(\boldsymbol w^{(n)}, \boldsymbol h_k )$ at point $\boldsymbol h_k=\boldsymbol h_k^{(n),(i)}$.
For small step size $\delta$, based on equation \eqref{applemmacon2eq1}, we can obtain a set of solutions $\boldsymbol h_k^{(n),(0)}$, $\boldsymbol h_k^{(n),(1)}$, $\boldsymbol h_k^{(n),(2)}$, $\cdots$, $\boldsymbol h_k^{(n),(i)}$, which satisfy,
\vspace{-.5em}
\begin{equation}\vspace{-.5em}
G_k(\boldsymbol w^{(n)}, \boldsymbol h_k^{(n),(0)})\geq 
  \cdots\geq G_k(\boldsymbol w^{(n)}, \boldsymbol h_k^{(n),(i)}).
\end{equation}
To provide the convergence condition for the gradient method, we introduce the definition of local accuracy, i.e., the solution $\boldsymbol h_k^{(n),(i)}$ of problem \eqref{sys0eq3_1} with accuracy $\eta$ means that:
\vspace{-.5em}
   \begin{align}\label{sys0eq3_6}\vspace{-.5em}
G_k &(\boldsymbol w^{(n)}, \boldsymbol h_k^{(n),(i)})-G_k(\boldsymbol w^{(n)}, \boldsymbol h_k^{(n)*})
\nonumber\\&
\leq  \eta (G_k (\boldsymbol w^{(n)}, \boldsymbol h_k^{(n),(0)})-G_k(\boldsymbol w^{(n)}, \boldsymbol h_k^{(n)*})),
 \end{align}
%
where $\boldsymbol h_k^{(n)*}$ is the optimal solution of problem \eqref{sys0eq3_1}.}
Each user is assumed to solve the {\color{myc5}local optimization problem} \eqref{sys0eq3_1} with a target accuracy $\eta$.
Next, in Lemma \ref{lemmacon2}, {\color{myc1}{we derive a lower bound on the number of local iterations needed to achieve a local accuracy $\eta$ in \eqref{sys0eq3_6}.}}

\begin{lemma}\label{lemmacon2}
Let $v=\frac{2}{(2-L\delta)\delta\gamma}$.
If we set step $\delta< \frac{2}{L}$ and run the gradient method
\vspace{-.5em}
\begin{equation}\label{sys1eq5_0}\vspace{-.5em}
i\geq v\log_2(1/\eta) 
\end{equation}
iterations 
at each user,
we can solve {\color{myc5}local optimization problem}  \eqref{sys0eq3_1} with an accuracy $\eta$.
\end{lemma}

\itshape {Proof:}  \upshape
See Appendix A.  \hfill $\Box$


The lower bounded derived in \eqref{sys1eq5_0} reflects the growing trend for the number of local iterations with respect to accuracy $\eta$. In the following, we use this lower bound to \emph{approximate the number of iterations} $I_k$ needed for local computations by each user.



%


In Algorithm 1, the iterative method involves
a number of global iterations (i.e., the value of $n$ in Algorithm 1) to achieve a global accuracy $\epsilon_0$ for the global FL model.
In other words, the solution $\boldsymbol w^{(n)}$ of problem \eqref{sys0eq2} with accuracy $\epsilon_0$  is a point such that
   \begin{align}\label{sys0eq3_7}
F  (\boldsymbol w^{(n)})-F(\boldsymbol w^*)\leq \epsilon_0 (F  (\boldsymbol w^{(0)})-F(\boldsymbol w^*)),
 \end{align}
where $\boldsymbol w^{*}$ is the actual optimal solution of problem \eqref{sys0eq2}.

To analyze the convergence rate of Algorithm~1, we make the following two assumptions on the loss function.
{\color{myc3}\begin{itemize}
  \item[-] A1: Function $F_k (\boldsymbol{w} )$ is $L$-Lipschitz, i.e., $\nabla^2 F_k (\boldsymbol{w} ) \preceq L \boldsymbol{I}$.
  \item[-]A2:  Function $F_k (\boldsymbol{w} )$ is $\gamma$-strongly convex, i.e., $\nabla^2 F_k (\boldsymbol{w} ) \succeq \gamma\boldsymbol{I}$.
\end{itemize}
{\color{myc3}The values of $\gamma$ and $L$ are determined by the loss function.}

These assumptions can be easily satisfied by widely used FL loss functions such as linear
or logistic loss functions \cite{8851249}.
}
%
%
Under assumptions A1 and A2, we provide the following theorem about {\color{myc3}convergence rate of  Algorithm~1 (i.e., the number of iterations needed for   Algorithm~1 to converge)}, where each user solves its {\color{myc5}local optimization problem} with a given accuracy.

\begin{theorem}
If we run Algorithm 1 with {\color{myc3}$0<\xi\leq \frac{\gamma}{L}$} for
\begin{equation}\label{sys1eq5_1}
n\geq\frac{a}{1-\eta}
\end{equation}
iterations with
$a=\frac{2L^2
}{\gamma^2\xi}\ln\frac{1}{\epsilon_0}$,
we have $F  (\boldsymbol w^{(n)})-F(\boldsymbol w^*)\leq \epsilon_0 (F  (\boldsymbol w^{(0)})-F(\boldsymbol w^*))$.
\end{theorem}

\itshape {Proof:}  \upshape
See Appendix B.  \hfill $\Box$

From Theorem 1, we observe that the number of global iterations $n$ increases with the local accuracy $\eta$  at the rate of $1/(1-\eta)$.
{\color{myc2}{From Theorem 1, we can also see that the FL performance depends on parameters $L$, $\gamma$, $\xi$, $\epsilon_0$  and $\eta$.
Note that the prior work in  \cite[Eq. (9)]{shamir2014communication} only studied the number of iterations needed for FL convergence under  the special case in which $\eta=0$.
Theorem~1 provides a general convergence rate for FL with an arbitrary  $\eta$.}}
{\color{myc3}Since the FL algorithm involves the accuracy of local computation and the result aggregation, it is hard to calculate the exact number of iterations needed for convergence.}
In the following, we use $\frac{a}{1-\eta}$ to approximate the number $I_0$ of global iterations.

{{{\color{myc5}
 For parameter setup, one way is to choose a very small value of $\xi$ , i.e., $\xi \approx0$ satisfying $0 < \xi \leq \frac{\gamma}{L}$, for an arbitrary learning task and loss function.
 However, based on  Theorem 1, the iteration time is pretty large for a very small  value of $\xi$. As a result, we first choose a large value of $\xi$ and decrease the value of $\xi$ to $\xi/2$ when the loss does not decrease over time (i.e., the value of $\xi$ is large). Note that in the simulations, the value of $\xi$ always changes at least three times. Thm 1 is suitable for the situation when the value of $\xi$ is fixed, i.e., after at most three iterations of Algorithm 1.
}}}

{\color{myc5}
\subsection{Extension to Nonconvex Loss Function}

We replace convex assumption A2 with the following condition:
 \begin{itemize}
  \item[-]B2:   Function $F_k (\boldsymbol{w} )$ is of $\gamma$-bounded nonconvexity (or $\gamma$-nonconvex), i.e., all the eigenvalues of $\nabla^2 F_k (\boldsymbol{w} ) $ lie in $[-\gamma, L]$, for some $\gamma\in(0,L]$.  
\end{itemize}
Due to the non-convexity of function $F_k (\boldsymbol{w} )$, we respectively replace $F_k (\boldsymbol{w} )$ and $F (\boldsymbol{w} )$ with their regularized versions \cite{allen2018natasha}
\begin{equation}\label{Regfun}
\tilde F_k^{(n)} (\boldsymbol{w} )=F_k (\boldsymbol{w} )+\gamma\|\boldsymbol w-\boldsymbol w^{(n)}\|^2, \tilde F^{(n)}(\boldsymbol w)=\sum_{k=1}^K \frac{D_k}{D} \tilde F_k^{(n)}(\boldsymbol w).
\end{equation}
Based on assuption A2 and \eqref{Regfun}, both functions $F_k^{(n)} (\boldsymbol{w} )$ and $F^{(n)} (\boldsymbol{w} )$ are $\gamma$-strongly convex, i.e., $\nabla^2 \tilde F_k^{(n)} (\boldsymbol{w} ) \succeq \gamma \boldsymbol{I}$ and $\nabla^2 \tilde F^{(n)} (\boldsymbol{w} ) \succeq \gamma \boldsymbol{I}$. Moreover, it can be proved  that  both functions $F_k^{(n)} (\boldsymbol{w} )$ and $F^{(n)} (\boldsymbol{w} )$ are $(L+2\gamma)$-Lipschitz.
The convergence analysis in Section II-B can be direcltly applied. }






\vspace{-1em}
\section{Resource Allocation for Energy Minimization}\vspace{-0.5em}

In this section, we formulate the energy minimization problem for FL.
Since it is challenging to obtain the globally optimal solution due to nonconvexity,
 an iterative algorithm with low complexity is proposed to solve the energy minimization problem.

\vspace{-.5em}
\subsection{Problem Formulation}\vspace{-.5em}
{\color{myc3}Our goal is to minimize the total energy consumption of all users under a latency constraint.
This energy efficient optimization problem can be posed as follows:
\vspace{-0.5em}
\begin{subequations}\label{sys1min1}\vspace{-.5em}
\begin{align}
\mathop{\min}_{\boldsymbol t, \boldsymbol b, \boldsymbol f, \boldsymbol p, \eta} \:&
 E,
\tag{\theequation}  \\
\textrm{s.t.} \quad\:\:\:
&\frac a {1-\eta}\left(  \frac{A_k\log_2(1/\eta)}{f_k} +t_k
\right)\leq T,\quad \forall k\in \mathcal K,\\
& t_kb_k\log_2\left(1+ \frac{g_kp_k}{N_0b_k}
\right) \geq s, \quad \forall k \in \mathcal K,\\
& \sum_{k=1}^K b_k\leq B, \\
& 0\leq f_k\leq f_k^{\max}, \quad \forall k \in \mathcal K,\\
& 0\leq p_k\leq p_k^{\max}, \quad \forall k \in \mathcal K,\\
& 0\leq \eta\leq 1,\\
& t_k\geq 0, b_k \geq 0, \quad \forall k \in \mathcal K,
\end{align}
\end{subequations}}
where $\boldsymbol t=[t_1, \cdots, t_K]^T$,
$\boldsymbol b=[b_1, \cdots, b_K]^T$,
$\boldsymbol f=[f_1, \cdots, f_K]^T$,
$\boldsymbol p=[p_1, \cdots, p_K]^T$,
{\color{myc5}$A_k=vC_k D_k$ is a constant},
 $f_k^{\max}$ and $p_k^{\max}$ are respectively the maximum local computation capacity and {\color{myc3}maximum value of the average transmit power} of user $k$.
{\color{myc5}Constraint  (\ref{sys1min1}a) is obtained by substituting $I_k=v\log_2(1/\eta)$ from \eqref{sys1eq5_0} and $I_0=\frac{a}{1-\eta}$ from \eqref{sys1eq5_1} into \eqref{Timeconseq1}. Constraints (\ref{sys1min1}b) is derived according to \eqref{sys1eq7} and $t_kr_k\geq s$.}
Constraint (\ref{sys1min1}a) indicates that the execution time of the local tasks and transmission time for all users should not exceed the maximum completion time for the whole FL algorithm.
{\color{myc3}Since the total number of iterations for each user is the same,
 the constraint in (\ref{sys1min1}a) captures a maximum time constraint for all users at each iteration if we divide the total number of iterations on both sides of  the constraint in (\ref{sys1min1}a).}
The data transmission constraint is given by (\ref{sys1min1}b),
while the bandwidth constraint is given by (\ref{sys1min1}c).
Constraints (\ref{sys1min1}d) and  (\ref{sys1min1}e) respectively represent the maximum local computation capacity and average transmit power limits of all users.
The local accuracy constraint is given by (\ref{sys1min1}f).

\vspace{-1em}
\subsection{Iterative Algorithm}\vspace{-0.5em}
The proposed iterative algorithm mainly contains two steps at each iteration.
To optimize $(\boldsymbol t, \boldsymbol b, \boldsymbol f, \boldsymbol p, \eta)$ in problem (\ref{sys1min1}), we first optimize $(\boldsymbol t, \eta)$ with fixed $(\boldsymbol b, \boldsymbol f, \boldsymbol p)$, then $(\boldsymbol b, \boldsymbol f, \boldsymbol p)$ is updated based on the obtained $(\boldsymbol t, \eta)$ in the previous step.
The advantage of this iterative algorithm lies in that we can obtain the optimal solution of $(\boldsymbol t, \eta)$ or $(\boldsymbol b, \boldsymbol f, \boldsymbol p)$ in each step.

In the first step, given $(\boldsymbol b, \boldsymbol f, \boldsymbol p)$, problem (\ref{sys1min1}) becomes:
\vspace{-0.5em}
\begin{subequations}\label{wte3min0}
\begin{align}
\mathop{\min}_{ \boldsymbol t, \eta} \quad&
\frac a {1-\eta} \sum_{k=1}^K \left(\kappa A_k\log_2(1/\eta) f_k^{2} + t_kp_k\right),
\tag{\theequation}  \\
\textrm{s.t.} \quad\:
&\frac a {1-\eta}\left(  \frac{A_k\log_2(1/\eta)}{f_k} +t_k
\right)\leq T,\quad \forall k\in \mathcal K,\\
& t_k  \geq t_k^{\min}, \quad \forall k \in \mathcal K,\\
& 0\leq \eta\leq 1,
\end{align}
\end{subequations}
where
\vspace{-0.5em}
\begin{equation}\vspace{-0.5em}
t_k^{\min} =\frac s {b_k\log_2\left(1+ \frac{g_kp_k}{N_0b_k}
\right)}, \quad \forall k\in\mathcal K.
\end{equation}

The optimal solution of \eqref{wte3min0} can be derived using
  the following theorem.

\begin{theorem}\label{IAtheorem2}
The optimal solution $(\boldsymbol t^*, \eta^*)$ of problem \eqref{wte3min0} satisfies:
\begin{equation}
t_k^*=t_k^{\min}, \quad \forall k\in\mathcal K,
\end{equation}
and $\eta^*$ is the optimal solution to:
\vspace{-0.5em}
\begin{subequations}\label{wte3min0_2}\vspace{-.5em}
\begin{align}
\mathop{\min}_{  \eta} \quad&
 \frac {\alpha_1\log_2(1/\eta)+ \alpha_2} {1-\eta}
\tag{\theequation}  \\
\textrm{s.t.} \quad\:
& \eta^{\min}\leq \eta\leq \eta^{\max},
\end{align}
\end{subequations}
where
\vspace{-0.5em}
\begin{equation}\vspace{-0.5em}
\eta^{\min}=\max_{k\in\mathcal K} \eta^{\min}_k,
\quad
\eta^{\max}=\min_{k\in\mathcal K} \eta^{\max}_k,
\end{equation}
$\beta_k(\eta_k^{\min})=\beta_k(\eta_k^{\max})=t_{k}^{\min}$, $t_k^{\min} \leq t_k^{\max}$, $\alpha_1$, $\alpha_2$ and $\beta_k(\eta)$ are defined in \eqref{wte3min0_1eq1}.
\end{theorem}

\itshape {Proof:}  \upshape
See Appendix C.  \hfill $\Box$

Theorem 2 shows that it is optimal to transmit with the  minimum time for each user.
Based on this finding, problem \eqref{wte3min0} is equivalent to the problem \eqref{wte3min0_2} with only one variable.
Obviously, the objective function \eqref{wte3min0_2} has a fractional form, which is generally hard to solve.
By using the parametric approach in \cite{dinkelbach1967nonlinear,yang2020energyefficient}, we consider the following problem,
\vspace{-0.5em}
\begin{equation}\label{equvialentProblem}\vspace{-0.5em}
H(\zeta)=\mathop{\min}_{\eta^{\min}\leq \eta\leq \eta^{\max}} {\alpha_1\log_2(1/\eta)+ \alpha_2} - \zeta{(1-\eta)}.
\end{equation}

It has been proved \cite{dinkelbach1967nonlinear} that solving (\ref{wte3min0_2}) is equivalent to finding the root of the nonlinear function $H(\zeta)$.
Since (\ref{equvialentProblem}) with fixed $\zeta$ is convex, the optimal solution $\eta^*$ can be obtained by setting the first-order derivative to zero, yielding the optimal solution:
$\eta^*=\frac{\alpha_1}{(\ln2)\zeta}$.
Thus, problem (\ref{wte3min0_2}) can be solved by using the Dinkelbach method in \cite{dinkelbach1967nonlinear} (shown as Algorithm~2).

\begin{algorithm}[t]
\caption{The Dinkelbach Method}
\label{alg:Framwork1}
\begin{algorithmic}[1]
\STATE Initialize $\zeta=\zeta^{(0)}>0$, iteration number $n=0$, and set the accuracy $\epsilon_3$.
\REPEAT
\STATE Calculate the optimal $\eta^*=\frac{\alpha_1}{(\ln2)\zeta^{(n)}}$  of problem (\ref{equvialentProblem}).
\STATE Update  $\zeta^{(n+1)}=\frac {\alpha_1\log_2(1/\eta^*)+ \alpha_2} {1-\eta^*} $
\STATE Set $n=n+1$.
\UNTIL $|H(\zeta^{(n+1)})|/|H(\zeta^{(n)})| < \epsilon_3$.
\end{algorithmic}
\end{algorithm}

In the second step, given $(\boldsymbol t, \eta)$ calculated in the first step, problem (\ref{sys1min1}) can be simplified as:
\begin{subequations}\label{wte3min00}\vspace{-.5em}
\begin{align}
\mathop{\min}_{ \boldsymbol b, \boldsymbol f, \boldsymbol p} \:&
 \frac {a  }{1-\eta} \sum_{k=1}^K \left(\kappa A_k\log_2(1/\eta) f_k^{2} + t_kp_k\right),
\tag{\theequation}  \\
\textrm{s.t.} \:\:\:
&\frac a {1-\eta}\left(  \frac{A_k\log_2(1/\eta)}{f_k} +t_k
\right)\leq T,\quad \forall k\in \mathcal K,\\
& t_kb_k\log_2\left(1+ \frac{g_kp_k}{N_0b_k}
\right) \geq s, \quad \forall k \in \mathcal K,\\
& \sum_{k=1}^K b_k\leq B, \\
& 0\leq p_k\leq p_k^{\max}, \quad \forall k \in \mathcal K,\\
& 0\leq f_k\leq f_k^{\max}, \quad \forall k \in \mathcal K.
\end{align}
\end{subequations}

Since both objective function and constraints can be decoupled, problem \eqref{wte3min00} can be decoupled into two subproblems:
\begin{subequations}\label{wte3min00_1}
\begin{align}
\mathop{\min}_{  \boldsymbol f} \quad&
 \frac {a\kappa\log_2(1/\eta)} {1-\eta} \sum_{k=1}^K     A_k  f_k^{2} ,
\tag{\theequation}  \\
\textrm{s.t.} \quad
&\frac a {1-\eta}\left(  \frac{A_k\log_2(1/\eta)}{f_k} +t_k
\right)\leq T,\quad \forall k\in \mathcal K,\\
& 0\leq f_k\leq f_k^{\max}, \quad \forall k \in \mathcal K,
\end{align}
\end{subequations}
and
\vspace{-0.5em}
\begin{subequations}\label{wte3min00_2}\vspace{-.5em}
\begin{align}
\mathop{\min}_{  \boldsymbol b,\boldsymbol p} \quad&
 \frac {a}{1-\eta} \sum_{k=1}^K  t_kp_k,
\tag{\theequation}  \\
\textrm{s.t.} \quad
& t_kb_k\log_2\left(1+ \frac{g_kp_k}{N_0b_k}
\right) \geq s, \quad \forall k \in \mathcal K,\\
& \sum_{k=1}^K b_k\leq B, \\
& 0\leq p_k\leq p_k^{\max}, \quad \forall k \in \mathcal K.
\end{align}
\end{subequations}

According to \eqref{wte3min00_1}, it is always efficient to utilize the minimum computation capacity $f_k$.
To minimize \eqref{wte3min00_1}, the optimal $f_k^*$ can be obtained from (\ref{wte3min00_1}a), which gives:
\vspace{-.5em}
\begin{equation}\label{wte3min00_1eq1}\vspace{-.5em}
f_k^*=\frac{aA_k\log_2(1/\eta)}
{ {T(1-\eta)}  -at_k }, \quad \forall k\in\mathcal K.
\end{equation}

%
%
We solve problem \eqref{wte3min00_2} using the following theorem.

\begin{theorem}\label{IAtheorem3_2}
The optimal solution $(\boldsymbol b^*, \boldsymbol p^*)$ of problem \eqref{wte3min00_2} satisfies:
\vspace{-0.5em}
\begin{equation}\label{wte3min00_21eq3}\vspace{-0.5em}
b_k^*=\max\{ b_k(\mu),b_k^{\min} \},
\end{equation}
and
\begin{equation}
p_k^*=\frac{N_0 b_k^*}{g_k}\left(2^{\frac{s}{ t_k b_k^*}}-1\right),
\end{equation}
where
\begin{equation}\label{wte3min00_2eq2_3}
  b_k^{\min}=-\frac{(\ln2) s}
{ t_k W\left(-\frac{(\ln2) N_0 s}{g_kp_k^{\max} t_k} \text e^{-\frac{(\ln2) N_0s }{g_kp_k^{\max} t_k}}
\right)+\frac {(\ln2)N_0s}{g_kp_k^{\max}}},
\end{equation}
$b_k(\mu)$ is the solution to
\begin{equation}\label{wte3min00_2eq2_5}
\frac{N_0}{g_kt_k}\left(e^{\frac{(\ln2)s}{ t_k b_k(\mu)}}-1-\frac{(\ln2)s}{ t_k b_k(\mu)} e^{\frac{(\ln 2)s}{ t_k b_k(\mu)}}\right) +\mu=0,
\end{equation}
and
$\mu$ satisfies
\begin{equation}\label{wte3min00_21eq5}
\sum_{k=1}^K\max\{ b_k(\mu),b_k^{\min} \}=B.
\end{equation}
\end{theorem}

\itshape {Proof:}  \upshape
See Appendix D.  \hfill $\Box$

\begin{algorithm}[t]
\caption{: Iterative Algorithm}
\begin{algorithmic}[1]
\STATE
 Initialize a feasible solution $( \boldsymbol t^{(0)}, \boldsymbol b^{(0)}, \boldsymbol f^{(0)}, \boldsymbol p^{(0)},$ $\eta^{(0)})$ of problem (\ref{sys1min1})  and set $l=0$.
 \REPEAT
\STATE With given $( \boldsymbol b^{(l)}, \boldsymbol f^{(l)}, \boldsymbol p^{(l)})$, obtain the optimal $(\boldsymbol t^{(l+1)}, \eta^{(l+1)})$ of problem \eqref{wte3min0}.
\STATE With given $(\boldsymbol t^{(l+1)}, \eta^{(l+1)})$, obtain the  optimal $($ $\boldsymbol b^{(l)},\boldsymbol f^{(l)}, \boldsymbol p^{(l)})$  of problem \eqref{wte3min00}.
\STATE Set $l=l+1$.
\UNTIL {objective value (\ref{sys1min1}a) converges}
\end{algorithmic}
\end{algorithm}

By iteratively solving problem \eqref{wte3min0} and  problem \eqref{wte3min00}, the algorithm that solves problem (\ref{sys1min1}) is given in Algorithm~3.
Since the optimal solution of   problem \eqref{wte3min0} or  \eqref{wte3min00} is obtained in each step, the objective value of  problem (\ref{sys1min1}) is nonincreasing in each step.
Moreover, the objective value of problem (\ref{sys1min1}) is lower bounded by zero.
Thus, Algorithm~3 always converges to a local optimal solution.


\subsection{Complexity Analysis}

To solve the energy minimization problem \eqref{sys1min1} by using Algorithm~3, the major complexity in each step lies in solving problem \eqref{wte3min0} and  problem \eqref{wte3min00}.
To solve problem \eqref{wte3min0}, the major complexity lies in obtaining the optimal $\eta^*$ according to Theorem 2, which involves complexity $\mathcal O(K\log2(1/\epsilon_1))$ with accuracy $\epsilon_1$ by using the Dinkelbach method.
To solve problem \eqref{wte3min00}, two subproblems \eqref{wte3min00_1} and \eqref{wte3min00_2} need to be optimized.
For subproblem \eqref{wte3min00_1}, the complexity is $\mathcal O(K)$ according to \eqref{wte3min00_1eq1}.
For subproblem \eqref{wte3min00_2}, the complexity is $\mathcal O(K\log_2(1/\epsilon_2)\log_2(1/\epsilon_3))$, where $\epsilon_2$ and $\epsilon_3$ are respectively the accuracy of solving \eqref{wte3min00_2eq2_3} and \eqref{wte3min00_2eq2_5} \cite{yang2020RIS}.
As a result, the total complexity of the proposed Algorithm 3 is $\mathcal O(L_{\text{it}}SK)$, where $L_{\text{it}}$ is the number of iterations for iteratively optimizing $(\boldsymbol t, \eta)$ and $(T, \boldsymbol b, \boldsymbol f, \boldsymbol p)$,
and $S= \log2(1/\epsilon_1)+\log2(1/\epsilon_2)\log2(1/\epsilon_3)$.

The conventional successive convex approximation (SCA) method can be used to solve problem \eqref{sys1min1}.
The complexity of SCA method is $O(L_{\text{sca}}K^3)$, where $L_{\text{sca}}$ is the total number of iterations for SCA method.
Compared to SCA, the proposed Algorithm 3 grows linearly with the number of users $K$.

It should be noted that Algorithm 3 is done at the BS side before executing the FL scheme in Algorithm 1.
{\color{myc2}{To implement Algorithm 3, the BS needs to gather the information of $g_k$, $p_k^{\max}$, $f_k^{\max}$, $C_k$, $D_k$, and $s$, which can be uploaded by all users before the FL process. Due to small data size, the transmission delay of these information can be neglected.}}
The BS broadcasts the obtained solution to all users.
Since the BS has high computation capacity, the latency of implementing Algorithm 3 at the BS will not affect the latency of the FL process.
\vspace{-.5em}
\section{Algorithm to Find a Feasible Solution of Problem \eqref{sys1min1}}\vspace{-.5em}
Since the iterative algorithm to solve problem \eqref{sys1min1} requires an initial feasible solution,
we provide an efficient algorithm to find a feasible solution of problem \eqref{sys1min1} in this section.
To obtain a feasible solution of \eqref{sys1min1}, we construct the completion time minimization problem:
\begin{subequations}\label{sys1min10}\vspace{-.5em}
\begin{align}
\mathop{\min}_{T, \boldsymbol t, \boldsymbol b, \boldsymbol f, \boldsymbol p, \eta} \:&
 T,
\tag{\theequation}  \\
\textrm{s.t.} \quad\:\: &(\ref{sys1min1}a)-(\ref{sys1min1}g).
\end{align}
\end{subequations}
We define $(T^*, \boldsymbol t^*, \boldsymbol b^*, \boldsymbol f^*, \boldsymbol p^*, \eta^*)$ as the optimal solution  of problem (\ref{sys1min10}).
Then, $(\boldsymbol t^*, \boldsymbol b^*, \boldsymbol f^*, \boldsymbol p^*, \eta^*)$ is a feasible solution of problem (\ref{sys1min1}) if $T^*\leq T$, where $T$ is the maximum delay in constraint (\ref{sys1min1}a). Otherwise, problem (\ref{sys1min1}) is infeasible.

Although the completion time minimization problem (\ref{sys1min10}) is still nonconvex due to constraints (\ref{sys1min1}a)-(\ref{sys1min1}b), we show that the globally optimal solution can be obtained by using the bisection method.
\vspace{-.5em}
\subsection{Optimal Resource Allocation}\vspace{-.5em}
\begin{lemma}\label{lemmatime1}
Problem (\ref{sys1min10}) with   $T<T^*$ does not have a feasible solution (i.e., it is infeasible), while  problem (\ref{sys1min10}) with  $T>T^*$ always has a feasible solution (i.e., it is feasible).
\end{lemma}


\itshape {Proof:}  \upshape
See Appendix E.  \hfill $\Box$

According to Lemma 2, we can use the bisection method to obtain the optimal solution of problem \eqref{sys1min10}.

With a fixed $T$, we still need to check whether there exists a feasible solution satisfying constraints (\ref{sys1min1}a)-(\ref{sys1min1}g).
From constraints (\ref{sys1min1}a) and (\ref{sys1min1}c), we can see that it is always efficient  to utilize the maximum computation capacity, i.e.,
$f_k^*=f_{k}^{\max},   \forall k \in \mathcal K$.
From (\ref{sys1min1}b) and (\ref{sys1min1}d), we can see that minimizing the completion time occurs when
$p_k^*=p_{k}^{\max}, \forall k \in \mathcal K$.
Substituting the maximum computation capacity and maximum transmission power into (\ref{sys1min10}), the completion time minimization problem  becomes:
\begin{subequations}\label{time2min2}\vspace{-.5em}
\begin{align}
 \min_{T, \boldsymbol t, \boldsymbol b, \eta} \quad\: &T
  \tag{\theequation}  \\
\textrm{s.t.} \quad\:
& t_k
 \leq \frac {(1-\eta)T}a +\frac{A_k\log_2\eta}{f_k^{\max}},\quad \forall k\in \mathcal K,\\
& \frac{s}{t_k} \leq b_k\log_2\left(1+ \frac{g_kp_k^{\max}}{N_0b_k}
\right), \quad \forall k \in \mathcal K,\\
& \sum_{k=1}^K b_k\leq B, \\
& 0\leq \eta\leq 1,\\
& t_k\geq 0, b_k \geq 0, \quad \forall k \in \mathcal K.
\end{align}
\end{subequations}

Next, we provide the sufficient and necessary condition for the feasibility of set (\ref{time2min2}a)-(\ref{time2min2}e).

\begin{lemma}\label{lemmatime11}
With a fixed $T$,  set (\ref{time2min2}a)-(\ref{time2min2}e) is nonempty if an only if
\vspace{-.5em}
\begin{equation}\label{time2min2eq1}\vspace{-.5em}
 B\geq \min_{0\leq \eta\leq 1} \quad \sum_{k=1}^K u_k(v_k(\eta)),
\end{equation}
where
\vspace{-.5em}
\begin{equation} \label{time2min2eq2}\vspace{-.5em}
u_k(\eta)=-\frac{(\ln2)  \eta }
{ W\left(-\frac{(\ln2) N_0 \eta }{g_kp_k^{\max}} \text e^{-\frac{(\ln2) N_0 \eta }{g_kp_k^{\max}}}
\right)+\frac {(\ln2) N_0\eta}{g_kp_k^{\max}}},
\end{equation}
and
\vspace{-.5em}
\begin{equation}\label{time2min2eq3}\vspace{-.5em}
v_k(\eta)=\frac{s}{\frac {(1-\eta)T}a +\frac{A_k\log_2\eta}{f_k^{\max}}}.
\end{equation}
\end{lemma}

\itshape {Proof:}  \upshape
See Appendix F.  \hfill $\Box$

To effectively solve \eqref{time2min2eq1} in Lemma \ref{lemmatime11}, we provide the following lemma.
\begin{lemma}\label{lemmatime2}
In \eqref{time2min2eq2}, $u_k(v_k(\eta))$ is a convex function.
\end{lemma}

\itshape {Proof:}  \upshape
See Appendix G.  \hfill $\Box$

Lemma \ref{lemmatime2} implies that the optimization problem in \eqref{time2min2eq1} is a convex problem, which can be effectively solved. 
By finding the optimal solution of \eqref{time2min2eq1},  the sufficient and necessary condition for the feasibility of set (\ref{time2min2}a)-(\ref{time2min2}e) can be simplified
using the following theorem. 

\begin{theorem}\label{theoremtime1}
Set (\ref{time2min2}a)-(\ref{time2min2}e) is nonempty if and only if
\vspace{-.5em}
\begin{equation}\label{time2min2eq5}\vspace{-.5em}
 B\geq \sum_{k=1}^K u_k(v_k(\eta^*)),
\end{equation}
where $\eta^*$ is the unique solution to
$\sum_{k=1}^K   u_k'(v_k(\eta^*))v_k'(\eta^*)=0$.
\end{theorem}

Theorem \ref{theoremtime1} directly follows  from Lemmas  \ref{lemmatime11} and \ref{lemmatime2}.
Due to the convexity of function $u_k(v_k(\eta))$, $\sum_{k=1}^K   u_k'(v_k(\eta^*))v_k'(\eta^*)$ is an increasing function of $\eta^*$. As a result, the unique solution of $\eta^*$ to $\sum_{k=1}^K   u_k'(v_k(\eta^*))v_k'(\eta^*)=0$ can be effectively solved via the bisection method.
Based on Theorem \ref{theoremtime1}, the algorithm for obtaining the minimum completion time is summarized in Algorithm~4.
{\color{myc1}{Theorem \ref{theoremtime1} shows that the optimal FL accuracy level $\eta^*$ meets the first-order condition $\sum_{k=1}^K   u_k'(v_k(\eta^*))v_k'(\eta^*)=0$, i.e., the optimal $\eta^*$ should not be too small or too large for FL.
This is because, for small $\eta$, the local computation time (number of iterations) becomes high as shown in  Lemma 1.
For large $\eta$, the transmission time is long due to the fact that a large number of global iterations is required as shown in Theorem 1.}}

\begin{algorithm}[t]
\caption{Completion Time Minimization}
\begin{algorithmic}[1]
 \STATE Initialize $T_{\min}$, $T_{\max}$, and the tolerance $\epsilon_5$.
 \REPEAT
 \STATE Set $T=\frac{T_{\min}+T_{\max}}{2}$.
 \STATE Check the feasibility condition (\ref{time2min2eq5}).
 \STATE If set (\ref{time2min2}a)-(\ref{time2min2}e) has  a feasible solution, set $T_{\max}=T$. Otherwise, set $T=T_{\min}$.
 \UNTIL $(T_{\max}-T_{\min})/T_{\max}\leq \epsilon_5$.
\end{algorithmic}
\end{algorithm}
\vspace{-.5em}
\subsection{Complexity Analysis}\vspace{-.5em}

The major complexity of Algorithm 4 at each iteration lies in checking the feasibility condition (\ref{time2min2eq5}).
To check the inequality in (\ref{time2min2eq5}), the optimal $\eta^*$ needs to be obtained 
by using the bisection method, which involves the complexity of $\mathcal O(K\log_2(1/\epsilon_4))$ with accuracy $\epsilon_4$.
As a result, the total complexity of Algorithm 4 is $\mathcal O(K\log_2(1/\epsilon_4)\log_2(1/\epsilon_5))$, where $\epsilon_5$ is the accuracy of the bisection method used in the outer layer.
{\color{myc1}{The complexity of  Algorithm 4 is low since $\mathcal O(K\log_2(1/\epsilon_4)\log_2(1/\epsilon_5))$ grows linearly with the total number of users.

Similar to Algorithm 3 in Section III, Algorithm 4 is done at the BS side before executing the FL scheme in Algorithm 1, which will not affect the latency of the FL process.}}

\vspace{-.5em}
\section{Numerical Results}\vspace{-.5em}
For our simulations, we deploy $K=50$ users uniformly in a square area of size $500$ m $\times$ $500$~m with the BS located at its center.
The path loss model is $128.1+37.6\log_{10} d$ ($d$ is in km)
and the standard deviation of shadow fading is $8$ dB.
In addition, the noise power spectral density is  $N_0=-174$ dBm/Hz.
{\color{myc3}We use the real open blog feedback dataset in \cite{buza2014feedback}.
This dataset with a total number of 60,000 data samples originates from blog posts and the dimensional of each data sample is 281.
The prediction task associated with the data is the prediction
of the number of comments in the upcoming 24 hours.}
Parameter $C_k$ is uniformly distributed in $[1,3]\times10^4$ cycles/sample.
{\color{myc3}The effective switched capacitance in local computation is $\kappa=10^{-28}$ \cite{7572018}.}
In Algorithm~1, we set  $\xi=1/10$, $\delta=1/10$, and $\epsilon_0=10^{-3}$. 
Unless specified otherwise, we choose an equal maximum average transmit power $p_1^{\max}=\cdots=p_K^{\max}=p^{\max}=10$ dB, an equal maximum computation capacity $f_1^{\max}=\cdots=f_K^{\max}=f^{\max}=2$ GHz, a transmit data size $s=28.1$ kbits,
 and a bandwidth $B=20$ MHz.
 Each user has $D_k=500$ data samples, which are randomly selected from the dataset with equal probability.
All statistical results are averaged over 1000 independent runs.

\vspace{-1em}
\subsection{Convergence Behavior}
\vspace{-.5em}

\begin{figure}
\centering
\includegraphics[width=3.0in]{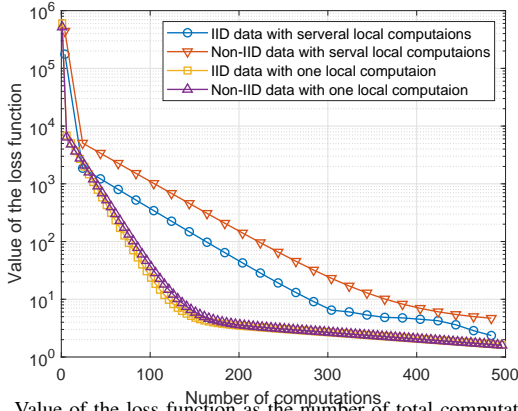}
\vspace{-1.5em}
\caption{{\color{myc5}Value of the loss function as the number of total computations varies for FL algorithms with IID and non-IID data.}}
\vspace{-1em}
\label{re3fig12}
\end{figure}
{\color{myc5}
Fig. \ref{re3fig12} shows the value of the loss function as the number of total computations varies with IID and non-IID data.
For IID data case, all data samples are first shuffled and then partitioned into
$60,000/500=120$ equal parts, and each device is assigned with one particular part.
For non-IID data case \cite{zhu2018low}, all
data samples are first sorted by digit label and then divided into 240 shards of size $250$, and
each device is assigned with two shards. Note that the latter is a pathological non-IID partition
way since most devices only obtain two kinds of digits.
From this figure, it is observed that the FL algorithm with non-IID data shows similar convergence behavior with the FL algorithms with IID data.
Besides, we find that the FL algorithm with multiple local updates requires more total computations than the FL algorithm with only one local update.

\begin{figure}
\centering
\includegraphics[width=3.0in]{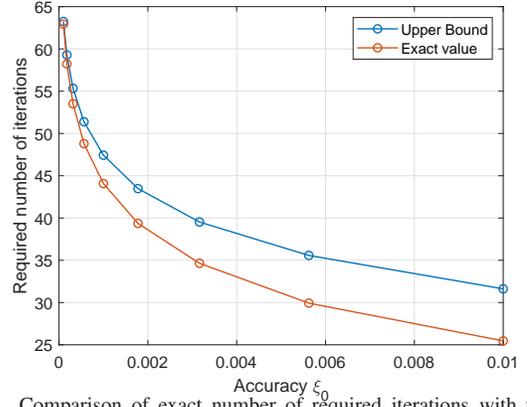}
\vspace{-1.5em}
\caption{Comparison of exact number of required iterations with the upper bound derived in Theorem 1.}
\vspace{-1em}
\label{re1fig1new}
\end{figure}

In Fig. \ref{re1fig1new}, we show the comparison of exact number of required iterations with the upper bound derived in Theorem 1. According to this figure, it is found that the gap of the exact number of required iterations with the upper bound derived in Theorem 1 decreases as the accuracy $\xi_0$ decreases, which indicates that the upper bound is tight for small value of $\xi_0$.} 

\vspace{-1em}
\subsection{Completion Time Minimization}
\vspace{-.5em}

\begin{figure}[t]
\centering
\includegraphics[width=3.0in]{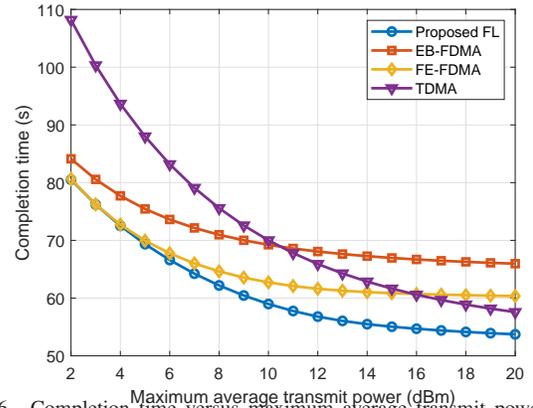}
\vspace{-1.5em}
\caption{Completion time  versus maximum average transmit power of each user.} \label{fig1}
\vspace{-1em}
\end{figure}

We compare the proposed FL scheme with the FL FDMA scheme with equal bandwidth $b_1=\cdots=b_K$ (labelled as `EB-FDMA'),
the FL FDMA scheme with fixed local accuracy $\eta=1/2$ (labelled as `FE-FDMA'), and the FL TDMA scheme in \cite{tran2019federated} (labelled as `TDMA').
Fig. \ref{fig1} shows how the completion time changes as the maximum average transmit power of each user varies.
We can see that the completion time of all schemes decreases with the maximum average transmit power of each user.
This is because a large maximum average transmit power can decrease the transmission time between users and the BS.
We can clearly see that the proposed FL scheme achieves the best performance among all schemes.
This is because the proposed approach jointly optimizes bandwidth and local accuracy $\eta$, while the bandwidth is fixed in EB-FDMA and $\eta$ is not optimized in FE-FDMA.
Compared to TDMA, {\color{myc1}{the proposed approach can reduce the completion time by up to 27.3\%}} due to the following two reasons.
First, each user can directly transmit result data to the BS after local computation in  FDMA, while the wireless transmission should be performed after the local computation for all users in TDMA, which needs a longer time compared to FDMA.
Second, the noise power for users in FDMA  is lower than in TDMA since each user is allocated to part of the bandwidth and each user occupies the whole bandwidth in TDMA, which indicates the transmission time in TDMA is longer than in FDMA.

{\color{myc5}
 Fig. \ref{fig5major} shows the completion time versus the maximum average transmit power with different batch sizes \cite{shallue2018measuring,lin2018don}.
The number of local iterations are keep fixed for these three schemes in Fig. \ref{fig5major}.
From this figure, it is found that the SGD method with smaller batch size (125 or 250) always outperforms the GD scheme, which indicates that it is efficient to use small batch size with SGD scheme.

\begin{figure}
\centering
\includegraphics[width=3.0in]{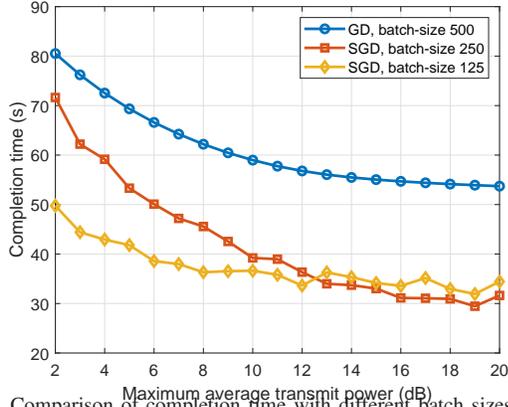}
\vspace{-1.5em}
\caption{Comparison of completion time with different batch sizes.} \label{fig5major}
\vspace{-1em}
\end{figure}
}

%
%

%
%

{\color{myc5}
 \begin{figure}[t]
\centering
\includegraphics[width=3.0in]{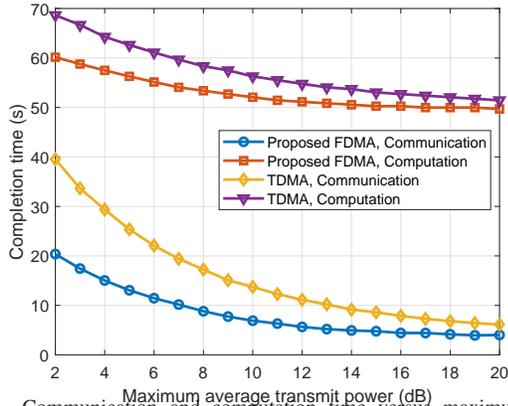}
\vspace{-1.5em}
\caption{Communication and computation time  versus maximum average transmit power of each user.} \label{fig2major}
\vspace{-1em}
\end{figure}
Fig. \ref{fig2major} shows how the communication and computation time change as the maximum average transmit power of each user varies.
It can be seen from this figure that both communication and computation time decrease with the maximum average transmit power of each user.
We can also see that the computation time is always larger than communication time and the decreasing speed of  communication time is faster than that of computation time.}

\vspace{-1em}
\subsection{Total Energy Minimization}
\vspace{-.5em}

\begin{figure}[t]
\centering
\includegraphics[width=3.0in]{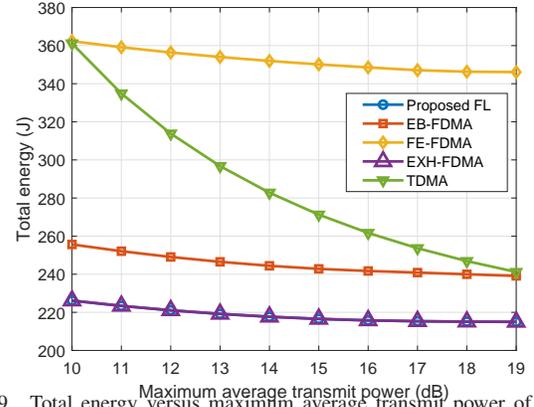}
\vspace{-1.5em}
\caption{Total energy versus maximum average transmit power of each user with $T=100$ s.} \label{fig6}
\vspace{-1em}
\end{figure}

Fig.~\ref{fig6} shows the  total energy as function of the maximum average transmit power of each user.
In this figure, the EXH-FDMA scheme is an exhaustive search method that can find a near optimal solution of
problem (\ref{sys1min1}), which refers to the proposed iterative algorithm with 1000 initial starting points.
There are 1000 solutions obtained by using EXH-FDMA, and the solution with the best objective value is treated as the near optimal solution.
From this figure, we can observe that the total energy decreases with the maximum average transmit power of each user.
Fig.~\ref{fig6} also shows that the proposed FL scheme outperforms the EB-FDMA, FE-FDMA, and TDMA schemes.
Moreover, the EXH-FDMA scheme achieves almost the same performance as the proposed FL scheme, which shows that the proposed approach achieves the optimum solution.

{\color{myc5}
\begin{figure}[t]
\centering
\includegraphics[width=3.0in]{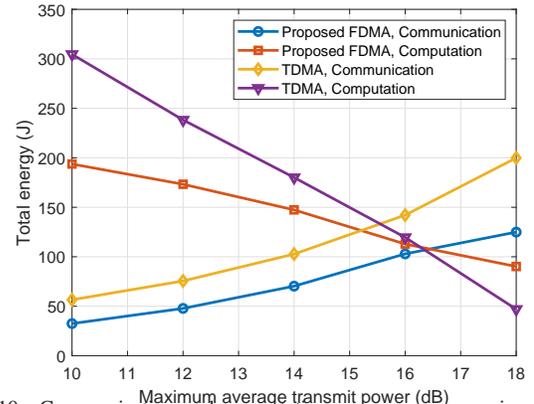}
\vspace{-1.5em}
\caption{Communication and computation energy  versus maximum average transmit power of each user.} \label{fig3major}
\vspace{-1em}
\end{figure}
Fig.~\ref{fig3major} shows the communication and computation energy as functions of the maximum average transmit power of each user.
In this figure, it is found that the communication energy increases with the  maximum average transmit power of each user, while the computation energy decreases with the maximum average transmit power of each user.
For low maximum average transmit power of each user (less than 15 dB), FDMA outperforms TDMA in terms of both computation and communication energy consumption.
For high maximum average transmit power of each user (higher than 17 dB), FDMA outperforms TDMA in terms of communication energy consumption, while TDMA outperforms FDMA in terms of computation energy consumption.}

\begin{figure}[t]
\centering
\includegraphics[width=3.0in]{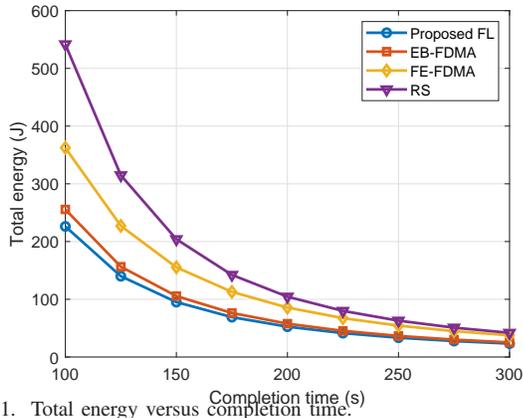}
\vspace{-1.5em}
\caption{Total energy versus completion time.} \label{fig10}
\vspace{-1.5em}
\end{figure}

Fig. \ref{fig10} shows the tradeoff between total energy consumption and completion time.
In this figure, we compare the proposed scheme with the random selection (RS) scheme, where 25 users are randomly selected out from $K=50$ users at each iteration.
We can see that FDMA outperforms RS in terms of total energy consumption especially for low completion time.
{\color{myc5}{This is because, in FDMA all users can transmit data to the BS in each global iteration, while only half number of users can transmit data to the BS in each global iteration. As a result, the average transmission time for each user in FDMA is larger than that in RS, which can lead to lower transmission energy for the proposed algorithm.}}
In particular, with given the same completion time,  the proposed FL can reduce energy of up to 12.8\%, 53.9\%, and  59.5\% compared to EB-FDMA, FE-FDMA, and RS, respectively.

\vspace{-.5em}
\section{Conclusions}
\vspace{-.5em}
In this paper, we have investigated the problem of energy efficient computation and transmission resource allocation of FL over wireless communication networks.
{\color{myc1}{We have derived the time and energy consumption models for FL based on the convergence rate.}}
With these models, we have formulated a joint learning and communication problem so as to
 minimize the total computation and transmission energy of the network.
To solve this problem, we have proposed an iterative algorithm with low complexity, for which, at each iteration, we have derived closed-form solutions for computation and transmission resources.
Numerical results have shown that the proposed scheme outperforms conventional schemes in terms of  total energy consumption, especially for small maximum average transmit power.

%

\appendices
{\vspace{-.5em}
\section{Proof of Lemma \ref{lemmacon2}}
\setcounter{equation}{0}
\renewcommand{\theequation}{\thesection.\arabic{equation}}
\vspace{-.5em}

Based on \eqref{applemmacon2eq1}, we have:
\vspace{-0.5em}
\begin{align}\label{applemmacon2eq2}\vspace{-0.5em}
&G_k(\boldsymbol w^{(n)}, \boldsymbol h_k^{(n),(i+1)})
 \overset{A1}{\leq} G_k(\boldsymbol w^{(n)}, \boldsymbol h_k^{(n),(i)})
\nonumber\\
&-\delta \left\|\nabla  G_k(\boldsymbol w^{(n)}, \boldsymbol h_k^{(n),(i)})\right\|^2
+\frac{L\delta^2}{2} \left\| \nabla  G_k(\boldsymbol w^{(n)}, \boldsymbol h_k^{(n),(i)})\right\|^2
\nonumber\\
&=G_k(\boldsymbol w^{(n)}, \boldsymbol h_k^{(n),(i)})
-\frac{(2-L\delta)\delta}{2} \left\|\nabla  G_k(\boldsymbol w^{(n)}, \boldsymbol h_k^{(n),(i)})\right\|^2.
\end{align}
Similar to \eqref{flconrproofeq1_2} in the following Appendix B, we can prove that:
\vspace{-0.5em}
\begin{align}\label{applemmacon2eq3}\vspace{-0.5em}
\|\nabla G_k(\boldsymbol w^{(n)}, \boldsymbol h_k)\|^2 \geq \gamma (G_k(\boldsymbol w^{(n)}, \boldsymbol h_k)-G_k(\boldsymbol w^{(n)}, \boldsymbol h_k^{(n)*})),
\end{align}
where $\boldsymbol h_k^{(n)*}$ is the optimal solution of problem \eqref{sys0eq3_1}.
Based on \eqref{applemmacon2eq2} and \eqref{applemmacon2eq3}, we have:
\vspace{-0.5em}
\begin{align}\label{applemmacon2eq5}\vspace{-0.5em}
&\quad G_k(\boldsymbol w^{(n)}, \boldsymbol h_k^{(n),(i+1)})-G_k(\boldsymbol w^{(n)}, \boldsymbol h_k^{(n)*})
\nonumber \\&\leq G_k(\boldsymbol w^{(n)}, \boldsymbol h_k^{(n),(i)})-G_k(\boldsymbol w^{(n)}, \boldsymbol h_k^{(n)*})
\nonumber \\&-\frac{(2-L\delta)\delta\gamma}{2}  (G_k(\boldsymbol w^{(n)}, \boldsymbol h_k^{(n),(i)})-G_k(\boldsymbol w^{(n)}, \boldsymbol h_k^{(n)*}))
\nonumber \\&
\leq\left(1
-\frac{(2-L\delta)\delta\gamma}{2}\right)^{i+1}
(G_k(\boldsymbol w^{(n)}, \boldsymbol 0)-G_k(\boldsymbol w^{(n)}, \boldsymbol h_k^{(n)*}))
   \nonumber\\&  {\leq}
\exp\left(
- (i+1)\frac{(2-L\delta)\delta\gamma}{2}\right)
  (G_k(\boldsymbol w^{(n)}, \boldsymbol 0)-G_k(\boldsymbol w^{(n)}, \boldsymbol h_k^{(n)*})),
\end{align}
where the last inequality follows from the fact that $1-x\leq \exp(-x)$.
To ensure that $G_k(\boldsymbol w^{(n)}, \boldsymbol h_k^{(n),(i)})-G_k(\boldsymbol w^{(n)}, \boldsymbol h_k^{(n)*})
\leq  \eta (G_k (\boldsymbol w^{(n)}, \boldsymbol 0)-G_k(\boldsymbol w^{(n)}, \boldsymbol h_k^{(n)*}))$, we have \eqref{sys1eq5_0}.}

\section{Proof of Theorem  1}
\setcounter{equation}{0}
\renewcommand{\theequation}{\thesection.\arabic{equation}}

Before proving Theorem 1, the following lemma is provided.
\begin{lemma}
Under the assumptions A1 and A2, the following conditions hold:
\vspace{-0.5em}
\begin{align}\label{flconrproofeq1}\vspace{-0.5em}
& \frac{1}{L}  \|\nabla  F_k(\boldsymbol w^{(n)}+\boldsymbol h^{(n)})-  \nabla  F_k(\boldsymbol w^{(n)})\|^2
\nonumber\\&
\leq(\nabla F_k(\boldsymbol w^{(n)}+\boldsymbol h^{(n)})-  \nabla F_k(\boldsymbol w^{(n)}))^T \boldsymbol h^{(n)}
\nonumber\\
\leq& \frac{1}{\gamma}  \|\nabla F_k(\boldsymbol w^{(n)}+\boldsymbol h^{(n)})-  \nabla F_k(\boldsymbol w^{(n)})\|^2,
\end{align}
and
\vspace{-0.5em}
\begin{equation}\label{flconrproofeq1_2}\vspace{-0.5em}
\|\nabla F(\boldsymbol w)\|^2 \geq \gamma (F(\boldsymbol w)-F(\boldsymbol w^*)).
\end{equation}
\end{lemma}

\itshape \textbf{Proof:}  \upshape
{\color{myc3}According to the Lagrange median theorem}, there always exists a $\boldsymbol w$ such that
\vspace{-0.5em}
\begin{equation}\label{flconrproofeq1_3}\vspace{-0.5em}
{  (\nabla F_k(\boldsymbol w^{(n)}+\boldsymbol h^{(n)})-  \nabla F_k(\boldsymbol w^{(n)}))}
 =\nabla^2 F_k(\boldsymbol w) { \boldsymbol h^{(n)}}.
\end{equation}
Combining assumptions A1, A2, and \eqref{flconrproofeq1_3} yields \eqref{flconrproofeq1}.

For the optimal solution $\boldsymbol w^*$ of  $F(\boldsymbol w^*)$, we always have $\nabla F(\boldsymbol w^{*})=\boldsymbol 0$.
Combining \eqref{sys0eq2} and \eqref{flconrproofeq1}, we also have $\gamma\boldsymbol{I} \preceq
\nabla^2 F (\boldsymbol{w} )$, which indicates that
\vspace{-0.5em}
\begin{equation}\label{flconrproofeq1_5}\vspace{-0.5em}
\|\nabla F(\boldsymbol w)-\nabla F(\boldsymbol w^*)\|
\geq \gamma \|\boldsymbol w-\boldsymbol w^* \|,
\end{equation}
and
{\color{myc3}
\vspace{-0.5em}
\begin{equation}\label{flconrproofeq1_6}\vspace{-0.5em}
F(\boldsymbol w^*) \geq
F(\boldsymbol w)+\nabla F(\boldsymbol w)^T   (\boldsymbol w^*-\boldsymbol w)+\frac {\gamma} 2 \|\boldsymbol w^*-\boldsymbol w\|^2 .
\end{equation}
}

As a result, we have:
\vspace{-0.5em}
\begin{align}\vspace{-0.5em}
\|\nabla F(\boldsymbol w)\|^2 &=\|\nabla F(\boldsymbol w)-\nabla F(\boldsymbol w^*)\|^2
\nonumber\\
&
\overset{\eqref{flconrproofeq1_5}}{\geq} \gamma \|\nabla F(\boldsymbol w)-\nabla F(\boldsymbol w^*)\|  \|\boldsymbol w-\boldsymbol w^* \|
\nonumber\\
& \geq \gamma (\nabla F(\boldsymbol w)-\nabla F(\boldsymbol w^*))^T   (\boldsymbol w-\boldsymbol w^* )
\nonumber\\
&
=\gamma  \nabla F(\boldsymbol w)^T   (\boldsymbol w-\boldsymbol w^*)
\nonumber\\
&
\overset{\eqref{flconrproofeq1_6}}{\geq} \gamma (F(\boldsymbol w)-F(\boldsymbol w^*)),
\end{align}
which proves \eqref{flconrproofeq1_2}.
 \hfill $\Box$

For the optimal solution of problem \eqref{sys0eq3_1}, the first-order derivative condition always holds, i.e.,
\begin{equation}\label{flconrproofeq2}
\nabla  F_k(\boldsymbol w^{(n)}+\boldsymbol h_k^{(n)*})-\nabla  F_k(\boldsymbol w^{(n)})+\xi\nabla  F(\boldsymbol w^{(n)}) =\boldsymbol 0.
\end{equation}

We are new ready to prove Theorem 1.
With the above inequalities and equalities, we have:
\vspace{-0.5em}
\begin{align}\label{flconrproofeq2_1}\vspace{-0.5em}
&F(\boldsymbol{w}^{(n+1)})
\nonumber\\&
 \overset{\eqref{sys0eq3_5},A1}{\leq} F(\boldsymbol{w}^{(n)})
+\frac {1} K \sum_{k=1}^K\nabla F(\boldsymbol{w}^{(n)})^T
\boldsymbol h_k^{(n)}
+\frac{L}{2K^2} \left\| \sum_{k=1}^K\boldsymbol h_k^{(n)}\right\|^2
\nonumber\\&
\overset{\eqref{sys0eq3_1}}= F(\boldsymbol{w}^{(n)})
+\frac {1} {K\xi} \sum_{k=1}^K\bigg[
G_k(\boldsymbol{w}^{(n)},\boldsymbol h_k^{(n)})
%
%
+ \nabla  F_k(\boldsymbol w^{(n)}) \boldsymbol h_k^{(n)}
\nonumber\\&\quad
- F_k(\boldsymbol{w}^{(n)}+\boldsymbol h_k^{(n)})
\bigg]
+\frac{L}{2K^2} \left\| \sum_{k=1}^K\boldsymbol h_k^{(n)}\right\|^2
\nonumber\\&\overset{A2}{\leq}
 F(\boldsymbol{w}^{(n)})
+\frac {1} {K\xi} \sum_{k=1}^K\bigg[
G_k(\boldsymbol{w}^{(n)},\boldsymbol h_k^{(n)})
- F_k(\boldsymbol{w}^{(n)})
\nonumber\\&\quad
-\frac {\gamma} 2 \|\boldsymbol h_k^{(n)}\|^2
\bigg]
+\frac{L}{2K^2} \left\| \sum_{k=1}^K\boldsymbol h_k^{(n)}\right\|^2.
\end{align}

According to the triangle inequality and mean inequality, we have
\vspace{-0.5em}
\begin{align}\label{flconrproofeq2_1_2}\vspace{-0.5em}
\left\|\frac{1}{K} \sum_{k=1}^K\boldsymbol h_k^{(n)}\right\|^2
\leq \left(\frac{1}{K} \sum_{k=1}^K \left\|\boldsymbol h_k^{(n)}\right\|\right)^2\leq \frac{1}{K} \sum_{k=1}^K \left\|\boldsymbol h_k^{(n)}\right\|^2.
\end{align}

Combining \eqref{flconrproofeq2_1} and \eqref{flconrproofeq2_1_2} yields:
\vspace{-0.5em}
\begin{align}\label{flconrproofeq2_2}\vspace{-0.5em}
&F(\boldsymbol{w}^{(n+1)})
\nonumber\\&
\leq
 F(\boldsymbol{w}^{(n)})
+\frac {1} {K\xi} \sum_{k=1}^K\bigg[
G_k(\boldsymbol{w}^{(n)},\boldsymbol h_k^{(n)})
- F_k(\boldsymbol{w}^{(n)})
\nonumber\\&\quad
-\frac {\gamma-L\xi} {2}\|\boldsymbol h_k^{(n)}\|^2
\bigg]
\nonumber\\&
\overset{\eqref{sys0eq3_1}}{=}
 F(\boldsymbol{w}^{(n)})
+\frac {1} {K\xi} \sum_{k=1}^K\bigg[
G_k(\boldsymbol{w}^{(n)},\boldsymbol h_k^{(n)})
- G_k(\boldsymbol{w}^{(n)},\boldsymbol h_k^{(n)*})
\nonumber\\&
\quad
-(G_k(\boldsymbol{w}^{(n)},\boldsymbol  0)
- G_k(\boldsymbol{w}^{(n)},\boldsymbol h_k^{(n)*}))
 -\frac {\gamma-L\xi} {2}\|\boldsymbol h_k^{(n)}\|^2
\bigg]
\nonumber\\&\overset{\eqref{sys0eq3_6}}{\leq}
 F(\boldsymbol{w}^{(n)})
-\frac {1} {K\xi} \sum_{k=1}^K\bigg[
  \frac {\gamma-L\xi} {2}\|\boldsymbol h_k^{(n)}\|^2
\nonumber\\&
\quad
 +(1-\eta)(G_k(\boldsymbol{w}^{(n)},\boldsymbol  0)
- G_k(\boldsymbol{w}^{(n)},\boldsymbol h_k^{(n)*}))
\bigg]
\nonumber\\&\overset{\eqref{sys0eq3_1}}{=}
 F(\boldsymbol{w}^{(n)})
-\frac {1} {K\xi} \sum_{k=1}^K\bigg[
  \frac {\gamma-L\xi} {2}\|\boldsymbol h_k^{(n)}\|^2
\nonumber\\&
\quad
 +(1-\eta)(F_k(\boldsymbol{w}^{(n)})- F_k(\boldsymbol w^{(n)}+ \boldsymbol h_k^{(n)*})
  \nonumber \\&\quad
+(\nabla  F_k(\boldsymbol w^{(n)})-  \xi\nabla  F(\boldsymbol w^{(n)}))^T \boldsymbol h_k^{(n)*})
\bigg]
\nonumber\\&\overset{\eqref{flconrproofeq2}}{=}
 F(\boldsymbol{w}^{(n)})
-\frac {1} {K\xi} \sum_{k=1}^K\bigg[
  \frac {\gamma-L\xi} {2}\|\boldsymbol h_k^{(n)}\|^2
\nonumber\\&
\quad
 +(1-\eta)(F_k(\boldsymbol{w}^{(n)})- F_k(\boldsymbol w^{(n)}+ \boldsymbol h_k^{(n)*})
  \nonumber \\&\quad
+\nabla  F_k(\boldsymbol w^{(n)}+\boldsymbol h_k^{(n)*})^T \boldsymbol h_k^{(n)*})
\bigg].
\end{align}

Based on assumption A2, we can obtain:
\vspace{-0.5em}
\begin{align}\label{flconrproofeq2_3}\vspace{-0.5em}
& F_k(\boldsymbol{w}^{(n)})
 \geq
 F_k(\boldsymbol w^{(n)}+ \boldsymbol h_k^{(n)*}) \nonumber\\&\quad
 +\nabla F_k(\boldsymbol w^{(n)}+ \boldsymbol h_k^{(n)*})^T (-\boldsymbol h_k^{(n)*})
 +\frac{\gamma}{2} \|\boldsymbol h_k^{(n)*} \|^2.
\end{align}
Applying \eqref{flconrproofeq2_3} to \eqref{flconrproofeq2_2}, we can obtain:
\vspace{-0.5em}
\begin{align}\label{flconrproofeq2_5}\vspace{-0.5em}
&F(\boldsymbol{w}^{(n+1)})\leq
 F(\boldsymbol{w}^{(n)})
-\frac {1} {K\xi} \sum_{k=1}^K\bigg[
  \frac {\gamma-L\xi} {2}\|\boldsymbol h_k^{(n)}\|^2
\nonumber\\&
\quad
 +
\frac{(1-\eta)\gamma}{2} \|\boldsymbol h_k^{(n)*} \|^2\bigg].
\end{align}

From \eqref{flconrproofeq1}, the following relationship is obtained:
\vspace{-0.5em}
\begin{equation}\label{flconrproofeq2_6_1}\vspace{-0.5em}
\|\boldsymbol h_k^{(n)*}\|^2
\geq \frac{1}{L^2} \|\nabla  F_k(\boldsymbol w^{(n)}+\boldsymbol h_k^{(n)*}))-\nabla F_k(\boldsymbol{w}^{(n)})\|^2.
\end{equation}
For the constant parameter $\xi$, we choose
\vspace{-0.5em}
\begin{equation}\label{flconrproofeq2_5_5}\vspace{-0.5em}
\gamma-L\xi>0.
\end{equation}

According to \eqref{flconrproofeq2_5_5} and \eqref{flconrproofeq2_5}, we can obtain:
\vspace{-.5em}
\begin{align}\label{flconrproofeq2_6}\vspace{-.5em}
F(\boldsymbol{w}^{(n+1)})&
{\leq}
 F(\boldsymbol{w}^{(n)})
-\frac {(1-\eta)\gamma} {2K\xi} \sum_{k=1}^K
 \|\boldsymbol h_k^{(n)*} \|^2
\nonumber\\& \overset{\eqref{flconrproofeq2_6_1}}{\leq}
 F(\boldsymbol{w}^{(n)})
-\frac {(1-\eta)\gamma} {2KL^2\xi} \sum_{k=1}^K\bigg[
\nonumber\\&
\quad
\|\nabla  F_k(\boldsymbol w^{(n)}+\boldsymbol h_k^{(n)*}) -\nabla F_k(\boldsymbol{w}^{(n)})\|^2 \bigg]
\nonumber\\&\overset{\eqref{flconrproofeq2}} {=}
 F(\boldsymbol{w}^{(n)})
- \frac {(1-\eta)\gamma\xi} {2L^2}  \|\nabla  F(\boldsymbol w^{(n)})\|^2
\nonumber\\&
\overset{\eqref{flconrproofeq1_2}} {\leq}
 F(\boldsymbol{w}^{(n)})
- \frac {(1-\eta)\gamma^2\xi} {2L^2}  (F(\boldsymbol w^{(n)})-F(\boldsymbol w^*)).
\end{align}

Based on \eqref{flconrproofeq2_6}, we get:
\vspace{-.5em}
\begin{align}\label{flconrproofeq2_7}\vspace{-.5em}
&F(\boldsymbol{w}^{(n+1)})-F(\boldsymbol w^*)
\nonumber\\
&  {\leq}
\left(1
- \frac {(1-\eta)\gamma^2\xi} {2L^2}\right)
  (F(\boldsymbol w^{(n)})-F(\boldsymbol w^*))
 \nonumber\\&
   {\leq}
\left(1
- \frac {(1-\eta)\gamma^2\xi} {2L^2}\right)^{n+1}
  (F(\boldsymbol w^{(0)})-F(\boldsymbol w^*))
   \nonumber\\&  {\leq}
\exp\left(
- (n+1)\frac {(1-\eta)\gamma^2\xi} {2L^2}\right)
  (F(\boldsymbol w^{(0)})-F(\boldsymbol w^*)),
\end{align}
where the last inequality follows from the fact that $1-x\leq \exp(-x)$.
To ensure that $F(\boldsymbol{w}^{(n+1)})-F(\boldsymbol w^*)\leq \epsilon_0(F  (\boldsymbol w^{(0)})-F(\boldsymbol w^*))$, we have \eqref{sys1eq5_1}.

\vspace{-1em}
\section{Proof of Theorem \ref{IAtheorem2}}
\setcounter{equation}{0}
\renewcommand{\theequation}{\thesection.\arabic{equation}}
\vspace{-.5em}

According to \eqref{wte3min0}, transmitting with minimal time is always energy efficient, i.e., the optimal time allocation is $t_k^*=t_k^{\min}$.
Substituting $t_k^*=t_k^{\min}$ into problem \eqref{wte3min0} yields:
\begin{subequations}\label{wte3min0_1}
\begin{align}
\mathop{\min}_{  \eta} \quad&
 \frac {\alpha_1\log_2(1/\eta)+ \alpha_2} {1-\eta}
\tag{\theequation}  \\
\textrm{s.t.} \quad\:
& t_k^{\min}
 \leq \beta_k(\eta),\quad \forall k\in \mathcal K,\\
& 0\leq \eta\leq 1,
\end{align}
\end{subequations}
where
\vspace{-.5em}
\begin{align}\label{wte3min0_1eq1}\vspace{-.5em}
\alpha_1=a\sum_{k=1}^K  \kappa A_kf_k^{2},
\alpha_2=a\sum_{k=1}^K t_k^{\min}p_k,
\nonumber
\\
\beta_k(\eta)=\frac {(1-\eta)T}a +\frac{A_k\log_2\eta}{f_k^{\max}}.
\end{align}


From \eqref{wte3min0_1eq1}, it can be verified that $\beta_k(\eta)$ is a concave function. 
Due to the concavity of $\beta_k(\eta)$, constraints (\ref{wte3min0_1}a) can be equivalently transformed to:
\vspace{-.5em}
\begin{equation}\label{wte3min0_1eq5}\vspace{-.5em}
\eta_k^{\min} \leq \eta \leq \eta_k^{\max},
\end{equation}
where $\beta_k(\eta_k^{\min})=\beta_k(\eta_k^{\max})=t_{k}^{\min}$ and $t_k^{\min} \leq t_k^{\max}$.
Since $\beta_k(1)=0$, $\lim_{\eta\rightarrow0+}\beta_k(\eta)=-\infty$ and $t_k^{\min}>0$, we always have $0<t_k^{\min} \leq t_k^{\max}<1$. 
With the help of \eqref{wte3min0_1eq5}, problem \eqref{wte3min0_1} can be
simplified as \eqref{wte3min0_2}.

\vspace{-.75em}
\section{Proof of Theorem \ref{IAtheorem3_2}}
\setcounter{equation}{0}
\renewcommand{\theequation}{\thesection.\arabic{equation}}
\vspace{-.5em}

To minimize $\sum_{k=1}^K t_kp_k$, transmit power $p_k$ needs to be minimized.
To minimize $p_k$ from (\ref{wte3min00_2}a), we have:
\vspace{-.5em}
\begin{equation}\label{wte3min00_2eq1}\vspace{-.5em}
p_k^*=\frac{N_0 b_k}{g_k}\left(2^{\frac{s}{ t_k b_k}}-1\right).
\end{equation}

The first order and second order derivatives of $p_k^*$ can be respectively given by
\vspace{-.5em}
\begin{equation}\label{wte3min00_2eq2}\vspace{-.5em}
\frac{\partial p_k^*}{\partial b_k}=\frac{N_0}{g_k}\left(e^{\frac{(\ln2)s}{ t_k b_k}}-1-\frac{(\ln2)s}{ t_k b_k} e^{\frac{(\ln 2)s}{ t_k b_k}}\right),
\end{equation}
and
\vspace{-.5em}
\begin{equation}\label{wte3min00_2eq2_1}\vspace{-.5em}
\frac{\partial^2 p_k^*}{\partial b_k^2}=\frac{N_0(\ln2)^2s^2}{g_kt_k^2 b_k^3}  e^{\frac{(\ln2)s}{ t_k b_k}} \geq 0.
\end{equation}
From \eqref{wte3min00_2eq2_1}, we can see that $\frac{\partial p_k^*}{\partial b_k}$ is an increasing function of $b_k$.
Since
$
\lim_{b_k\rightarrow 0+}\frac{\partial p_k^*}{\partial b_k}=0,
$
we have $\frac{\partial p_k^*}{\partial b_k}<0$ for $0<b_k<\infty$, i.e., $p_k^*$ in \eqref{wte3min00_2eq1} is a decreasing function of $b_k$.
Thus, maximum transmit power constraint $p_k^*\leq p_k^{\max}$ is equivalent to:
\vspace{-.5em}
\begin{equation}\vspace{-.5em}
b_k\geq b_k^{\min}\triangleq-\frac{(\ln2) s}
{ t_k W\left(-\frac{(\ln2) N_0 s}{g_kp_k^{\max} t_k} \text e^{-\frac{(\ln2)N_0s }{g_kp_k^{\max} t_k}}
\right)+\frac {(\ln2) N_0s}{g_kp_k^{\max}}},
\end{equation}
where $b_k^{\min}$ is defined in \eqref{wte3min00_2eq2_3}.

Substituting \eqref{wte3min00_2eq1} into problem \eqref{wte3min00_2}, we can obtain:
\begin{subequations}\label{wte3min00_21}\vspace{-.5em}
\begin{align}
\mathop{\min}_{   \boldsymbol b} \quad&
  \sum_{k=1}^K\frac{N_0 t_k b_k }{g_k}\left(2^{\frac{s}{b_k t_k}}-1\right),
\tag{\theequation}  \\
\textrm{s.t.} \quad
& \sum_{k=1}^K b_k \leq B, \\
&b_k\geq b_k^{\min}, \quad \forall k \in \mathcal K,
\end{align}
\end{subequations}

According to \eqref{wte3min00_2eq2_1}, problem \eqref{wte3min00_21} is a convex function, which can be effectively solved by using the Karush-Kuhn-Tucker conditions.
The Lagrange function of \eqref{wte3min00_21} is:
\vspace{-.5em}
\begin{equation}\label{wte3min00_21eq1}\vspace{-.5em}
\mathcal L (\boldsymbol b, \mu)= \sum_{k=1}^K\frac{N_0 t_k b_k }{g_k}\left(2^{\frac{s}{b_k t_k}}-1\right)+\mu\left( \sum_{k=1}^K b_k- B
\right),
\end{equation}
where $\mu$ is the Lagrange multiplier associated with constraint (\ref{wte3min00_21}a).
The first order derivative of $\mathcal L (\boldsymbol b, \mu)$ with respect to $b_k$ is:
\vspace{-.5em}
\begin{equation}\label{wte3min00_21eq2}\vspace{-.5em}
\frac{\partial\mathcal L (\boldsymbol b, \mu)}{\partial b_k}= \frac{N_0}{g_kt_k}\left(e^{\frac{(\ln2)s}{ t_k b_k}}-1-\frac{(\ln2)s}{ t_k b_k} e^{\frac{(\ln 2)s}{ t_k b_k}}\right) +\mu.
\end{equation}
We define $ b_k(\mu)$ as the unique solution to $\frac{\partial\mathcal L (\boldsymbol b, \mu)}{\partial b_k}=0$.
Given constraint (\ref{wte3min00_21}b), the optimal $b_k^*$ can be founded from \eqref{wte3min00_21eq3}.
Since the objective function \eqref{wte3min00_21} is a decreasing function of $b_k$, constrain (\ref{wte3min00_21}a) always holds with equality for the optimal solution, which shows that the optimal Lagrange multiplier is obtained by solving \eqref{wte3min00_21eq5}.

\vspace{-.5em}
\section{Proof of Lemma \ref{lemmatime1}}
\setcounter{equation}{0}
\renewcommand{\theequation}{\thesection.\arabic{equation}}
\vspace{-.5em}

Assume that problem (\ref{sys1min10}) with  $T=\bar T<T^*$ is feasible, and the feasible solution is $(\bar T, \bar{ \boldsymbol t}, \bar{\boldsymbol b}, \bar{\boldsymbol f}, \bar{\boldsymbol p},\bar\eta)$.
Then, the solution $(\bar T, \bar{ \boldsymbol t}, \bar{\boldsymbol b}, \bar{\boldsymbol f}, \bar{\boldsymbol p}, \bar\eta)$ is feasible with lower value of the objective function than solution $(T^*, \boldsymbol t^*, \boldsymbol b^*, \boldsymbol f^*, \boldsymbol p^*,\eta^*)$, which contradicts the fact that $(T^*, \boldsymbol t^*, \boldsymbol b^*, \boldsymbol f^*, \boldsymbol p^*,\eta^*)$ is the optimal solution.

For problem (\ref{sys1min10}) with  $T=\bar T>T^*$, we can always construct a feasible solution $(\bar T, \boldsymbol t^*, \boldsymbol b^*, \boldsymbol f^*, \boldsymbol p^*, \eta^* )$ to problem (\ref{sys1min10})   by checking all constraints.

\vspace{-.5em}
\section{Proof of Lemma \ref{lemmatime11}}
\setcounter{equation}{0}
\renewcommand{\theequation}{\thesection.\arabic{equation}}
\vspace{-.5em}

To prove this, we first define function
\vspace{-.5em}
\begin{equation}\label{apptime2min2eq5_0}\vspace{-.5em}
y=x \ln\left( 1+ \frac{1}{x}
\right), \quad x>0.
\end{equation}
Then, we have
\vspace{-.5em}
\begin{equation}\label{apptime2min2eq5_1}\vspace{-.5em}
 y'=\ln\left( 1+ \frac{1}{x}
\right) -\frac{1}{x+1},
 y''= -\frac{1}{x(x+1)^2}<0.
\end{equation}

According to  \eqref{apptime2min2eq5_1}, $y'$ is a  decreasing function.
Since $\lim_{t_i \rightarrow +\infty} y' =0$, we can obtain that $y'>0$ for all $0<x<+\infty$.
As a result, $y$ is an increasing function, i.e., the right hand side of (\ref{time2min2}b) is an increasing function of bandwidth $b_k$.

To ensure that the maximum bandwidth constraint (\ref{time2min2}c) can be satisfied, the left hand side of (\ref{time2min2}b) should be as small as possible, i.e., $t_k$ should be as long as possible.
Based on  (\ref{time2min2}a), the optimal time allocation should be:
\vspace{-.5em}
\begin{equation}\label{apptime2min2eq5_3}\vspace{-.5em}
t_k^*=\frac {(1-\eta)T}a +\frac{A_k\log_2\eta}{f_k^{\max}},\quad \forall k\in \mathcal K.
\end{equation}

Substituting \eqref{apptime2min2eq5_3} into (\ref{time2min2}b), we can construct  the following problem:
\begin{subequations}\label{apptime2min5}\vspace{-.5em}
\begin{align}
 \min_{\boldsymbol b, \eta} \quad\: & \sum_{k=1}^K b_k
  \tag{\theequation}  \\
\textrm{s.t.} \quad\:
& v_k(\eta)\leq b_k\log_2\left(1+ \frac{g_kp_k^{\max}}{N_0b_k}
\right),\quad \forall k \in \mathcal K,\\
& 0\leq \eta\leq 1,\\
& b_k \geq 0, \quad \forall k \in \mathcal K,
\end{align}
\end{subequations}
where $v_k(\eta)$  is defined in \eqref{time2min2eq3}.
We can observe that set (\ref{time2min2}a)-(\ref{time2min2}e) is nonempty if an only if the optimal objective value of \eqref{apptime2min5} is less than $B$.
Since the right hand side of (\ref{time2min2}b) is an increasing function,
(\ref{time2min2}b) should hold with equality for the optimal solution of problem \eqref{apptime2min5}.
Setting  (\ref{time2min2}b) with equality, problem \eqref{apptime2min5} reduces to \eqref{time2min2eq1}.

\vspace{-1em}
\section{Proof of Lemma \ref{lemmatime2}}
\setcounter{equation}{0}
\renewcommand{\theequation}{\thesection.\arabic{equation}}
\vspace{-.5em}

We first prove that $v_k(\eta)$ is a convex function.
To show this, we define:
\vspace{-.5em}
\begin{equation}\label{apptime2min2eq1}\vspace{-.5em}
\phi(\eta)=\frac{s}{\eta},\quad 0\leq \eta\leq1,
\end{equation}
and
\vspace{-.5em}
\begin{equation}\label{apptime2min2eq2}\vspace{-.5em}
\varphi_k(\eta)={\frac {(1-\eta)T}a +\frac{A_k\log_2\eta}{f_k^{\max}}}, \quad 0\leq \eta\leq1.
\end{equation}
According to \eqref{time2min2eq3}, we have:
$
v_k(\eta)=\phi(\varphi_k(\eta)).
$
Then, the second-order derivative of $v_k(\eta)$ is:
\vspace{-0.5em}
\begin{align}\label{apptime2min2eq3}\vspace{-0.5em}
v_k''(\eta)&=\phi''(\varphi_k(\eta))(\varphi_k'(\eta))^2+\phi'(\varphi_k(\eta))\varphi_k''(\eta).
\end{align}
According to \eqref{apptime2min2eq1} and \eqref{apptime2min2eq2}, we have:
\vspace{-.5em}
\begin{equation}\label{apptime2min2eq1_1}\vspace{-.5em}
\phi'(\eta)=-\frac{s}{\eta^2} \leq 0, \quad \phi''(\eta)=\frac{2s}{\eta^3} \geq 0,
\end{equation}
\vspace{-.5em}
\begin{equation}\label{apptime2min2eq2_1}\vspace{-.5em}
\varphi_k''(\eta)= -\frac{A_k}{(\ln2)f_k^{\max}\eta^2}  \leq 0.
\end{equation}
Combining \eqref{apptime2min2eq3}-\eqref{apptime2min2eq2_1}, we can find that $v_k''(\eta)\geq 0$, i.e., $v_k(\eta)$ is a convex function.

Then, we can show that $u_k(\eta)$ is an increasing and convex function.
According to Appendix B, $u_k(\eta)$ is the inverse function of the right hand side of (\ref{time2min2}b).
If we further define function:
\vspace{-.5em}
\begin{equation}\label{apple2time2min2eq5_1}\vspace{-.5em}
z_k(\eta)=\eta\log_2\left(1+ \frac{g_kp_k^{\max}}{N_0\eta}
\right), \quad \eta \geq 0,
\end{equation}
$u_k(\eta)$ is the inverse function of $z_k(\eta)$, which gives
$u_k(z_k(\eta))=\eta$.

According to (\ref{apptime2min2eq5_0}) and(\ref{apptime2min2eq5_1}) in Appendix B, function $z_k(\eta)$ is an increasing and concave function, i.e., $z_k'(\eta)\geq 0$ and $z_k''(\eta)\leq 0$.
Since $z_k(\eta)$ is an increasing function, its inverse function $u_k(\eta)$ is also an increasing function.

Based on the definition of concave function, for any $\eta_1\geq0$, $\eta_2\geq 0$ and $0\leq\theta\leq1$, we have:
\vspace{-.5em}
\begin{equation}\label{apple2time2min2eq5_3}\vspace{-.5em}
z_k(\theta\eta_1+(1-\theta)\eta_2) \geq \theta z_k(\eta_1)+
(1-\theta)z_k(\eta_2).
\end{equation}
Applying the increasing  function $u_k(\eta)$ on both sides of \eqref{apple2time2min2eq5_3} yields:
\vspace{-.5em}
\begin{equation}\label{apple2time2min2eq5_5}\vspace{-.5em}
 \theta\eta_1+(1-\theta)\eta_2  \geq u_k(\theta z_k(\eta_1)+
(1-\theta)z_k(\eta_2)).
\end{equation}
Denote $\bar \eta_1=z_k(\eta_1)$ and $\bar \eta_2=z_k(\eta_2)$, i.e.,
we have $ \eta_1=u_k(\bar\eta_1)$ and $ \eta_2=u_k(\bar\eta_2)$.
Thus, \eqref{apple2time2min2eq5_5} can be rewritten as:
\vspace{-.5em}
\begin{equation}\label{apple2time2min2eq5_6}\vspace{-.5em}
\theta u_k(\bar\eta_1)+(1-\theta)u_k(\bar\eta_1)  \geq u_k(\theta \bar\eta_1 +
(1-\theta) \bar \eta_2 ),
\end{equation}
which indicates that $u_k(\eta)$ is a convex function.
As a result, we have proven that $u_k(\eta)$ is an increasing and convex function, which shows:
\vspace{-.5em}
\begin{equation}\label{apple2time2min2eq5_7}\vspace{-.5em}
u_k'(\eta)\geq 0, \quad u_k''(\eta)\geq 0.
\end{equation}

To show the convexity of $u_k(v_k(\eta))$, we have:
\vspace{-.5em}
\begin{align}\label{apptime2min2eq6}\vspace{-.5em}
u_k''(v_k(\eta))=u_k''(v_k(\eta))(v_k'(\eta))^2+u_k'(v_k(\eta))v_k''(\eta) \geq 0,
\end{align}
according to $v_k''(\eta)\geq 0$ and \eqref{apple2time2min2eq5_7}.
As a result, $u_k(v_k(\eta))$ is a convex function.

\vspace{-3em}
\bibliographystyle{IEEEtran}
\bibliography{IEEEabrv,ref}

\end{document}